\documentclass{emulateapj}

\usepackage{amsmath}
\usepackage{amssymb}
\usepackage{graphicx}

\shortauthors{Roth, N. and Kasen, D. }
\shorttitle{Implicit Monte Carlo Radiation-Hydrodynamics}

\begin{document}

\title{Monte Carlo Radiation-Hydrodynamics with implicit methods}
\slugcomment{ Accepted to ApJS } 

\author{Nathaniel Roth\altaffilmark{1}}
\author{Daniel Kasen\altaffilmark{1,2,3}}
\email{nathaniel.roth@berkeley.edu}
\altaffiltext{1}{Physics Department, University of California, Berkeley, CA 94720, USA}
\altaffiltext{2}{Astronomy Department and Theoretical Astrophysics Center, University of California, Berkeley, CA 94720, USA}
\altaffiltext{3}{Nuclear Science Division, Lawrence Berkeley National Laboratory, Berkeley, CA 94720, USA}

\begin{abstract} 
We explore the application of Monte Carlo transport methods to solving coupled radiation-hydrodynamics problems.
We use a time-dependent, frequency-dependent, 3-dimensional radiation transport code that is  special relativistic and includes some detailed microphysical interactions such as resonant line scattering. We couple the transport code to two different 1-dimensional (non-relativistic) hydrodynamics solvers: a spherical Lagrangian scheme and a Eulerian Godunov solver. The gas-radiation energy coupling is treated implicitly, allowing us to take hydrodynamical time-steps that are much longer than the radiative cooling time. We validate the code and assess its performance using a suite of radiation hydrodynamical test problems, including ones in the radiation energy dominated regime. We also develop techniques that reduce the noise of the Monte Carlo estimated radiation force by using the spatial divergence of the radiation pressure tensor. The results suggest that Monte Carlo techniques hold promise for simulating the multi-dimensional radiation hydrodynamics of astrophysical systems. 
\end{abstract}


\section{Introduction}

The dynamical effects of radiation can be important in astrophysical contexts, so numerical simulations must often address the radiation transport problem. The radiation field, when treated fully, is a function of not only three spatial coordinates, but also of time, frequency and two direction angles. The high dimensionality of the problem makes it computationally very challenging, and approximate methods that ignore certain dependencies (e.g., on frequency and/or angle) are often employed. Recent efforts aim to relax these approximations and improve the accuracy of the transport scheme. Given the difficulty of the radiation-hydrodynamics (RHD) problem, and the critical importance of it in astrophysical simulation, a number of different numerical techniques should be explored. Ultimately, no single approach may prove ideal in every conceivable application, and the relevant tradeoffs in performance will need to be considered on a case by case basis.

In this paper we explore the  coupling of Monte Carlo radiative transfer (MCRT) to hydrodynamics. The Monte Carlo approach offer several advantages as compared to a deterministic solution of the radiative transfer equation. MCRT generalizes readily to arbitrary 3-dimensional geometries, and can naturally incorporate multi-frequency, multi-angle, and time-dependent transport effects. It is also straightforward to include  complex physical interactions, such as anisotropic and inelastic scattering processes, polarization, and resonant line scattering. MCRT methods generally parallelize well (although not necessarily trivially for memory intensive  problems \citep{Kasen2008}) and so can be run profitably on massively-parallel machines. This last consideration may ultimately prove to be the most significant, as the available computing power increases over time.
 
The main disadvantage of MCRT methods is  the presence of stochastic error, such that the computation of a large number of  packet trajectories may be required. A number of variance reduction techniques exist to help limit the unwanted effects of noise, and certain acceleration techniques can alleviate the well known computational inefficiency of MCRT in regimes of high optical depth. The ultimate  expense of MCRT relative to other transport methods is difficult to estimate, but generally as the dimensionality of the problem increases, the advantages of Monte Carlo methods become more apparent. This suggests that MCRT will be competitive in addressing the full 3-D multi-angle multi-frequency RHD problem.

Here we present calculations using a MCRT code designed to handle the full-dimensionality of the Boltzmann transport problem -- i.e.,  the dependence on 3 spatial dimensions, time, frequency and angle. The code is special relativistic and includes some more complex physical interactions, such as resonant line scattering. It makes use of implicit techniques \citep{Fleck1971} in order to permit time-steps larger than the gas-radiation energy coupling time. For the sake of demonstrating the essential principles and assessing the viability of the approach, we restrict ourselves to coupling to a one-dimensional hydrodynamics solver; upcoming work will generalize to multi-dimensional RHD.

In section \ref{Sec:LiteratureReview}, we review some of the existing literature on RHD in astrophysics, including previous efforts in MCRT. In section \ref{Sec:Equations} we outline the equations solved and the simplifying assumptions employed. Section \ref{Sec:MonteCarlo} describes the Monte Carlo implementation, while section \ref{Sec:Hydro} describes our numerical hydrodynamics scheme. Section \ref{Sec:ImplicitMC} describes the implicit Monte Carlo technique and its use in our code. Section \ref{Sec:RadiationTests} describes some radiation-only tests of our frequency-dependent transfer code. Section \ref{Sec:RadhydroTests}, the centerpiece of this paper, presents a suite of RHD test problems. Section \ref{Sec:RadiationForceComparsion} shows how Monte Carlo noise can be reduced by computing the radiation force via spatial derivatives of the Eddington tensor, rather than through a direct Monte Carlo force estimator. Section \ref{Sec:Performance} provides some brief considerations of the numerical performance of our code and the possibilities for improving it in the future. Finally, Section \ref{Sec:Conclusion} presents our conclusions.

\section{Existing astrophysical radiation-hydrodynamics techniques}
\label{Sec:LiteratureReview}

Radiation-hydrodynamics is a vast topic that spans many scientific disciplines. In this brief (and necessarily incomplete) review, we will emphasize multi-purpose astrophysical fluid codes.

One of the oldest and most commonly used techniques is flux-limited diffusion (FLD) \citep{Levermore1981, Swesty2009}. As its name suggests, in this approach radiation is transported via a diffusion equation , which amounts to dropping all terms in the radiative transfer (RT) equation with a higher-order than linear angular dependence. An interpolation procedure connects the optically thick to optically thin regimes and ensures that the transfer rate of radiative energy never exceeds the speed of light. Grid-based hydrodynamic and magneto-hydrodynamic (MHD) codes in use today making use of FLD include those described in \citet{Turner2001, Hayes2006}, \citet{Krumholz2007}, \citet{Gittings2008}, \citet{Swesty2009}, \citet {Commercon2011}, \citet{van-der-Holst2011}, \citet{Orban2013}, \citet{Tomida2013}, \citet{Zhang2013},  \citet{Bryan2014}, \citet{Kolb2013}, and \citet{DAngelo2013}. Additionally, \citet{Whitehouse2004} describe a smoothed-particle hydrodynamics code that makes use of FLD.

While fast and relatively easy to implement, FLD suffers from some well-characterized shortcomings. It restricts the radiative flux to be in the direction of the radiative energy gradient, which can lead to misdirected radiation forces. As a result, the radiation in an FLD simulation will wrap around opaque barriers rather than cast sharp shadows.
  
An alternative method which alleviates this problem is the $M_1$ closure for the Eddington tensor \citep{Dubroca1999}. Here, the two lowest-order angular moments of the RT equation are used. The radiation energy and pressure are related via an entropy minimization procedure, which results in the correct behavior in the free-streaming and diffusion limits. The $M_1$ closure has been implemented in astrophysical RHD codes including those described in \citet{Gonzalez2007}, \citet{Aubert2008}, \citet{Vaytet2011}, \citet{Skinner2013}, \citet{Rosdahl2013}, \citet{Sadowski2014}, and \citet{McKinney2014}.

Another option is to solve the full RT equation for discretized solid angle bins. This can be accomplished by solving the equation along rays that extend through multiple cells in the domain, a technique referred to as ray-tracing or a long-characteristics method. An early example of an astrophysical RHD code to use this approach is the stellar atmoshpere code described in \citet{Nordlund1982}, \citet{Nordlund1990}, and \citet{Stein1998}, using a variation of a method first proposed by \citet{Feautrier1964}. The long characteristics approach is especially effective in situations where a single or small number of luminous point sources are present, and a common application is tracking the ionizing radiation emitted from massive stars. \citet{Abel2002} describes an an adaptively branching ray tracing procedure, and it has been applied to HD and MHD calculations coupled to ionizing radiation as described in \citet{Sokasian2003}, \citet{Krumholz2007-2} and \citet{Wise2011}. Other uses of ray tracing to track ionizing radiation include those described in \citet{Whalen2006}, \citet{Alvarez2006}, \citet{Henney2009}, and smoothed-particle hydrodynamics implementations are described in \citet{Gritschneder2009} and \citet{Greif2009}. \citet{Kuiper2010} introduced a hybrid approach in which ray tracing is used to follow the direct radiation from a point source, while FLD is used simultaneously to follow diffuse radiation.

A related method is the short-characteristics technique, which is a subset of $S_N$ methods. Here, the RT equation is solved for a fixed set of angles within every grid cell. Early descriptions of such a technique were given by \citet{Mihalas1978} and \citet{Kunasz1988}. For problems in which the radiation enters the diffusion regime, so that the radiative emissivity is distributed over many grid cells, the short-characteristics approach allows the the computational expense of the problem to scale more slowly with the size of the grid than in the long characteristic case \citep{Davis2012}. An early example of an astrophysical RHD code to make use of the short-characteristics technique for two-dimensional problems is described in \citet{Stone1992}. \citet{Liebendorfer2004}, \citet{Livne2004}, \citet{Buras2006}, and \citet{Ott2008} describe codes that use short characteristics for neutrino transport, which is coupled to hydrodynamics in simulations of core-collapse supernovae. \citet{Vogler2005} describe an RHD code that uses short-characteristics in the context of stellar atmospheres. \citet{Rijkhorst2006} developed a hybrid method that combined techniques from both short- and long-characeterstics radiative transfer codes in the context of adaptive mesh refinement in three dimensions. Three-dimensional MHD simulations with radiation tracked using short characteristics include those described in \citet{Heinemann2007}, \citet{Hayek2010}, and \citet{Jiang2012}. Overall, the short characteristics approach has proven to be accurate and reasonably fast. One potential drawback is the appearance of ray artifacts at large distances from luminous sources.

\citet{Petkova2011} introduced an advection technique to solve the monochromatic radiative transfer equation on both structured and unstructured meshes. In the most general implementation of this scheme, the flux of radiative energy between zones is discretized into solid angle cones, which bears some resemblance to the short-characteristics method.

MCRT, while used for decades to simulate spectra and light curves of astrophysical objects
\citep[e.g.][]{Mazzali1993, Kasen2006, Kerzendorf2014}, has only recently been employed in the context of astrophysical RHD. \citet{Lucy2005} developed time-dependent (non-stationary) MCRT techniques for outflows in which radiation was not dynamically important. \citet{Ercolano2011} used MCRT to process snapshots of HD simulations to demonstrate that the diffuse radiation field in stellar ionization problems can differ significantly from an on-the-spot approximation for remitted ionizing photons. \citet{Haworth2012} moved beyond snapshots and coupled MCRT of ionizing radiation, including the diffuse radiation field, to a hydrodynamics solver. \citet{Abdikamalov2012} applied MCRT to neutrino transport in core-collapse supernova simulations, and introduced extensions to the implicit Monte Carlo technique to the case of velocity-dependent transfer case (see section \ref{Sec:ImplicitMC}). \citet{Ghosh2011} and \citet{Garain2012} used MCRT to simulate the effects of Compton cooling in black hole accretion, and coupled this to a hydrodynamics solver. \citet{Noebauer2012} presented a general-purpose code that couples MCRT with a Godunov solver for hydrodynamics, and validated its performance in a suite of common RHD test problems. \citet{Wollaeger2013} combined implicit Monte Carlo with discrete diffusion techniques in high velocity outflows on a Lagrangian grid. 

In this work, we proceed in a manner similar to \citet{Noebauer2012}, and repeat some of the test problems contained therein. We keep our radiation equations exact to all orders in $v/c$ (see sections \ref{Sec:Equations} and \ref{Sec:MonteCarlo}), although our hydrodynamics equations remain non-relativistic. Unlike previous studies of RHD using Monte Carlo, we include test problems in which the radiation energy is dominant over thermal energy, and where the radiation pressure is dynamically important. We compare two techniques that may be used to calculate the force from radiation pressure. Our approach also makes use of the implicit Monte Carlo technique.

\section{Equations solved and simplifying assumptions}
\label{Sec:Equations}
We review here a basic formulation of radiation-hydrodynamics. For this, we rely heavily on the expositions in \citet{Mihalas1984} and \citet{Mihalas2001}, quoting directly many of the equations therein for ease of reference throughout the rest of this paper.

The equations governing the fluid flow are the mass conservation equation, the gas momentum conservation equation and the gas total (kinetic plus thermal) energy conservation equation, with source terms relating to radiative transfer. To order $v/c$, where $v$ is the fluid velocity and $c$ the speed of light, the equations are \citep{Mihalas2001}
\begin{equation}
\frac{\partial \rho}{dt} + \frac{\partial( \rho v^i)}{\partial x^i} = 0 \, \, ,
\label{Eq:MassConservation}
\end{equation}

\begin{equation}
\frac{\partial( \rho v^i)}{dt} + \frac{\partial}{\partial x^j} \left( \rho v^i v^j + p_0 \delta^{ij} \right)  = \rho f^i + G^i -\frac{v^i}{c} G^0 \, \, ,
\label{Eq:MomentumConservation}
\end{equation}

\begin{eqnarray}
\frac{\partial}{dt}\left[ \rho \left( \frac{1}{2}v^2 + e_0 \right) \right] &+& \frac{\partial}{\partial x^i} \left\{ \left[ \rho \left(\frac{1}{2} v^2 + e_0 \right) + p_0 \right] v^i \right\} \nonumber \\ &=& \rho v^i f^i + cG^0 \, \, .
\label{Eq:EnergyConservation}
\end{eqnarray}
We have used the Einstein summation convention for indices, and the generic superscript index $i$ may refer to the $x$, $y$, or $z$ component of a vector in a Cartesian coordinate system. The subscript $0$ denotes that quantities that are evaluated in the local comoving frame of the fluid. Otherwise, the quantity is evaluated in the frame of the fixed coordinate system, which we refer to as the lab frame. Thus, $\rho$ is the lab frame fluid density, $p_0$ is the comoving gas pressure, $e_0$ is the comoving gas specific internal energy (units energy per mass), and $f^i$ is a body force such as gravity as measured in the lab frame. The quantities $G^0$ and $G^i$ are the lab frame components of the force four-vector, ${\bf G}$. This four-vector specifies the energy and momentum coupling between the fluid and the radiation, and  will be defined explicitly below.

An ideal gas equation of state relates the comoving pressure and specific internal energy of the gas
\begin{equation}
p_0 = (\gamma_{\rm ad} - 1) \rho e_0 \, \, .
\end{equation}

We do not consider here the fully special relativistic fluid equations. However, in our treatment of the radiation transport we will be careful to include all special relativistic terms. We also have not included terms for viscous transport, thermal heat conduction, or an internal energy source terms such as would arise in a fluid undergoing nuclear reactions, although these can in principle be included as well \citep{Mihalas2001}.

To find the the radiation force four-vector ${\bf G}$, we begin with the lab frame radiative transfer equation
\begin{equation}
\frac{1}{c} \frac{\partial I_{\nu}(\bf n)}{\partial t} + n^i\frac{\partial I_{\nu}(\bf n)}{\partial x^i} = - \chi_{\nu} ( {\bf n}) I_{\nu}( {\bf n}) + \eta_{\nu}({\bf n}) \, \, .
\label{Eq:RadiativeTransfer}
\end{equation}
Here $I_{\nu}$ is the specific intensity of the radiation, $\nu$ is the frequency, $\chi_{\nu}$ (units cm$^{-1}$) is the total extinction coefficient, $\eta$ is the total radiative emissivity, and ${\bf n}$ is a unit vector representing a direction. We may also make reference to the radiative source function $S_{\nu} \equiv \eta_{\nu} / \chi_{\nu}$. For notational brevity we will henceforth suppress writing the dependence of $I_{\nu}$, $\chi_{\nu}$ and  $\eta_{\nu}$ on direction ${\bf n}$, and keep in mind that $I_{\nu}$, $\chi_{\nu}$ and $\eta_{\nu}$ are also be functions of position and time. Both $\chi_{\nu}$ and $\eta_{\nu}$ have contributions from scattering as well as thermal absorption and re-emission, as we will discuss below.

It is useful to define moments of the radiation intensity which correspond to radiative energy density, flux, and pressure
\begin{equation}
E_{\nu} =  \frac{1}{c}  \oint I_{\nu} \, d\Omega, \qquad E = \int_0^{\infty} E_{\nu} d{\nu} \, \, ,
\label{Eq:RadEnergyDefinition}
\end{equation}
\begin{equation}
F_{\nu}^i =  \oint I_{\nu} n^i \, d\Omega, \qquad F^i = \int_0^{\infty} F_{\nu}^i d{\nu} \, \, ,
\label{Eq:Eq:RadFluxDefinition}
\end{equation}
\begin{equation}
P_{\nu}^{ij} = \frac{1}{c}   \oint I n^i n^j \, d\Omega, \qquad P^{ij} = \int_0^{\infty} P_{\nu}^{ij} d{\nu} \, \, .
\label{Eq:PressureTensorDefinition}
\end{equation}

Equation \ref{Eq:RadiativeTransfer} may be integrated over frequency and solid angle to obtain the radiation energy equation
\begin{equation}
\frac{\partial E}{dt} + \frac{\partial F^i}{\partial x^i} = \int_0^{\infty} d{\nu} \oint d\Omega \left( -\chi_{\nu} I_{\nu} + \eta_{\nu} \right)   \equiv - c G^0 \, \, .
\label{Eq:G0Definition}
\end{equation}
This is a conservation equation for the radiation energy density. The integral represents an energy loss term for the radiation field, and hence an energy source term for the fluid, and can therefore be identified with $-c G^0$.

Integrating equation \ref{Eq:RadiativeTransfer} over frequency and then integrating with respect to $n^i d\Omega$ results in the radiation momentum equation
\begin{equation}
\frac{1}{c^2}\frac{\partial F^i}{dt} + \frac{\partial P^{ij}}{\partial x^j} = \frac{1}{c} \int_0^{\infty} d{\nu} \oint d\Omega \left[ \left(-\chi_{\nu} I_{\nu} + \eta_{\nu} \right ) n^i\right]  \equiv -G^i \, \, .
\label{Eq:GiDefinition}
\end{equation}
This is a conservation equation for the radiation momentum density, and the integral can be identified with the  term $-c G^i$. The problem has now been fully posed up to the specification of initial conditions and boundary conditions for the radiation and the fluid. 

At this point, we will introduce some simplifying assumptions that will allow us to derive relatively simple expressions for the radiation four-force. These  remain in effect for the entirety of this paper, although some or all of them could be relaxed in future work: (1) All absorption and emission, including scattering processes, are isotropic in the comoving frame. (2) Scattering in the comoving frame is elastic (energetically coherent). (3) The quantities  $\eta$ and $\chi$ can be decomposed into separate thermal and scattering contributions
\begin{eqnarray}
\chi_{0 \nu} &=& \chi_{0 \nu}^t + \chi_{0 \nu}^s \, \, , \nonumber \\
\eta_{0 \nu} &=& \eta_{0 \nu}^t + \eta_{0 \nu}^s \, \, .
\end{eqnarray}
We will sometimes refer to $\chi_{0 \nu}^t$ as the absorption coefficient. We will also find it useful to define an opacity\footnote{Our notation here differs slightly from \citet{Mihalas2001}, in which the symbol $\kappa_0$ is used for the extinction coefficient (units cm$^{-1}$) rather than for an opacity (units cm$^{-2}$ g$^{-1}$).}   (units cm$^2$ g$^{-1}$) as  $\kappa_{0 \nu} = \chi_{0 \nu}/\rho_0$, and $\kappa_{0 \nu}^t = \chi_{0\nu}^t/\rho_0$, $\kappa_{0 \nu}^s = \chi_{0 \nu}^s/\rho$. We define an opacity ratio 
\begin{equation}
\epsilon_{\nu} \equiv \frac{\kappa_{0 \nu}^t}{\kappa_{0 \nu}} = \frac{\chi_{0 \nu}^t}{\chi_{0 \nu}} \, \, .
\label{Eq:EpsilonDefinition}
\end{equation}
The case $\epsilon_{\nu} = 0$ corresponds to complete scattering of photons, without any thermal absorption or re-emission. The case $\epsilon_{\nu} = 1$ corresponds to a situation with no scattering, in which every photon interaction corresponds to a photon being absorbed and its energy transferred to the gas.

For the thermal emission, we assume local thermodynamic equilibrium (LTE). In this case Kirchoff's law implies that the thermal component of the emissivity, $\eta_{0 \nu}^t$, is equal to $\chi^t_{0 \nu}B_{0 \nu}$, where $B_{0 \nu}$ is the Planck function calculated using the gas temperature measured in the comoving frame. Then we may write the total (thermal plus scattering) emissivity in the comoving frame as 
\begin{eqnarray}
\eta_{0\nu} &=& \chi_{0\nu} \left[ \epsilon_{\nu} B_{0 \nu} + (1 - \epsilon_{\nu}) \frac{c E_{0\nu}}{4 \pi} \right] \, \, .
\end{eqnarray}
When this expression for $\eta_{0\nu}$ is plugged into the comoving frame analogue of Equation~\ref{Eq:G0Definition}, the scattering out of the beam ($\chi^s_{0 \nu} I_{0\nu}$) cancels the scattering into the beam ($\eta^s_{0 \nu}$) , and the energy component of ${\bf G_0}$ becomes
\begin{equation}
\label{Eq:G0Intermediate}
c G^0_0 = \int_0^{\infty} \epsilon_{\nu} \chi_{0\nu} \left(c E_{0 \nu} - 4 \pi B_{0 \nu} \right) d \nu \, \, .
\end{equation}
Meanwhile, the assumed isotropy of emitted radiation, both thermal and scattering, allows us to simplify the spatial components of the force four-vector,
\begin{equation}
G^i_0 = \frac{1}{c} \int_0^{\infty} \chi_{0\nu} F^i_{0 \nu} d \nu \, \, .
\end{equation}
Finally, we introduce three mean extinction coefficients (energy-weighted mean, Planck mean, and flux-weighted mean, respectively),
\begin{eqnarray}
\chi_{0E} &=& \frac{\int_0^{\infty} \epsilon_\nu \chi_{0\nu} E_{0\nu}d \nu}{\int_0^{\infty} E_{0\nu} d\nu} \, \, ,\\
\chi_{0P} &=& \frac{\int_0^{\infty} \epsilon_\nu \chi_{0\nu} B_{0\nu}d \nu}{\int_0^{\infty} B_{0\nu} d\nu}\, \, ,\\
\chi_{0F} &=& \frac{\int_0^{\infty} \chi_{0\nu} F_{0\nu}d \nu}{\int_0^{\infty} F_{0\nu} d\nu} \, \, .
\end{eqnarray}
The expressions for the components of ${\bf G_0}$ then reduce to
\begin{eqnarray}
\label{Eq:ComovingG0}
c G^0_0 &=& c\left(\chi_{0E} E_0 - \chi_{0P}  a_r T_{0,g}^4 \right )  \, \, ,\\
\label{Eq:ComovingGi}
G^i_0 &=& \chi_{0F} F_0^i /c \, \, ,
\end{eqnarray}
where $a_r = 7.5657 \times 10^{-15}$ erg cm$^{-3}$ Kelvin$^{-4}$ is the radiation constant, which arises from the integration of the Planck function over frequency. 

We may then use a Lorentz transformation to determine the components of ${\bf G}$ in the lab frame
\begin{eqnarray}
\label{Eq:G0Simplified}
&G^0& = \gamma \left[G^0_0 + \left(\frac{v^i}{c}\right)G_0^i \right] \, \, ,
\\
&G^i& = G_0^i + \gamma \frac{v^i}{c} \left[ G_0^0 + \frac{\gamma}{\gamma + 1}\left(\frac{v^j}{c}\right)G_0^j \right] \, \, ,
\label{Eq:GiSimplified}
\end{eqnarray}
where $\gamma \equiv (1 - v^iv^i/c^2)^{-1/2}$. Equations \ref{Eq:G0Simplified} and \ref{Eq:GiSimplified} are accurate to all orders of $(v/c)$ \citep{Mihalas2001}. These two equations, along with the fluid equations (\ref{Eq:MassConservation} through \ref{Eq:EnergyConservation}), provide the high-level schematic for what our code solves. It is important to recognize that these are mixed-frame equations in the sense that the left-hand side refers to  a lab frame quantity, while the right-hand side is written in terms of comoving quantities.

We still must specify the expressions we will use to compute the comoving quantities $E_0$ and $F^i_0$. One approach would be to first construct the radiation energy density, flux and pressure entirely in the lab frame, and then relate the lab and comoving values of these quantities using the fact that they are components of a second rank Lorentz covariant tensor, the radiation stress-energy tensor. This approach is described in \citet{Mihalas2001}, and it is also the way to derive the equations used in \citet{Lowrie1999} and \citet{Jiang2012}, which have been truncated at order $(v/c)^2$. However, as will become clear below and in section \ref{Sec:MonteCarlo}, we are able to easily construct estimators of the flux in the comoving frame, and so we have a means of accurately calculating ${\bf G}$ without needing to compute and store the components of the pressure tensor.

We note that the radiation field, $I_\nu$, is ultimately composed of photons with four-momenta given by ${\bf M} = (h\nu/c) (1,n^i)$. The lab frame and comoving frame components of the four-momentum are related via a Lorentz transformation
\begin{equation}
\frac{\nu_0}{\nu} = \gamma(1 - n^i v^i/c) \, \, ,
\label{Eq:EnergyDoppler}
\end{equation}
and
\begin{equation}
n^i_0 = \left( \frac{\nu_0}{\nu} \right )^{-1} \left[ n^i -\frac{\gamma v^i}{c}
\left( 1 - \frac{\gamma n^j v^j/c}{\gamma +1} \right) \right] \, \, .
\label{Eq:dtrans}
\end{equation}
These equations incorporate the relevant Doppler shift and aberration effects. Two final transformations we will need are \citep{Thomas1930}
\begin{equation}
I_{0\nu} = \left(\frac{\nu_0}{\nu}\right)^3 I_\nu \, \, ,
\end{equation}
\begin{equation}
d\nu_0 d\Omega_0 = \left(\frac{\nu_0}{\nu}\right)^{-1} d\nu d\Omega \, \, .
\end{equation}

Then we may write the comoving radiation energy density and flux as
\begin{equation}
E_0 = \frac{1}{c} \int_0^{\infty} d\nu_0 \oint{I_0 d\Omega_0} = \frac{1}{c}\int_0^{\infty} d\nu \oint{ I_\nu \left(\frac{\nu_0}{\nu}\right)^2  d\Omega } \, \, ,
\label{Eq:ComovingEnergy}
\end{equation}
\begin{equation}
F^i_0 = \int_0^{\infty} d\nu_0 \oint{ I_0 n^i_0 d\Omega_0 } = \int_0^{\infty} d\nu \oint{ I_\nu \left(\frac{\nu_0}{\nu}\right)^2 n^i_0 d\Omega } \, \, .
\label{Eq:ComovingFlux}
\end{equation}

Equations \ref{Eq:ComovingEnergy} and \ref{Eq:ComovingFlux} provide us with a means of computing $E_0$ and $F_0$, accurate to all orders in $v/c$, in terms of integrals of lab frame quantities (with the help of equations \ref{Eq:EnergyDoppler} and \ref{Eq:dtrans}). It is straightforward to show that these equations are equivalent to those that follow from the Lorentz covariance of the stress-energy tensor \citep{Mihalas1984}.

Finally, we consider an approximate alternative to equation \ref{Eq:GiSimplified} that is valid in the raditaive diffusion regime. Consider once again the convservation of radiation momentum as expressed in equation \ref{Eq:GiDefinition}. As noted in \citet{Mihalas2001}, when the radiation is diffusing, the time-derivative on the left-hand side of that equation is at most on the order $\lambda_p / l$ compared to the radiation force term on the right-hand side of the equation, where $\lambda_p$ is the photon mean free path and $l$ is the fluid flow length scale. Since $\lambda_p / l \ll 1$ by definition in the radiative diffusion regime, this term can be safely dropped, leaving us with
\begin{eqnarray}
\label{Eq:PressureDivergence}
G^i &=& -\frac{\partial P^{ij}}{\partial x^j} \, \, , \qquad {\rm or} \nonumber \\
G^i &=& -\frac{\partial \left(f^{ij} E \right) }{\partial x^j} \, \, , 
\end{eqnarray}
where in the second line we have introduced the Eddington tensor $f^{ij}$ which satisfies
\begin{equation}
P^{ij} = f^{ij}E \, \, .
\end{equation}
In some situations, using equation \ref{Eq:PressureDivergence} for $G^i$ may reduce the Monte Carlo sampling noise (see section \ref{Sec:RadiationForceComparsion}). 

\section{Monte Carlo Transport}
\label{Sec:MonteCarlo}

To make use of equations \ref{Eq:ComovingEnergy} and either \ref{Eq:ComovingFlux} or \ref{Eq:PressureDivergence}, we still must solve the radiation transfer equation for $I$. In the MCRT approach, one forgoes a direct numerical solution in favor of a stochastic simulation of photon transport. The radiation field is represented by discrete packets which are tracked through randomized scatterings and absorptions. Each packet is described by  a lab frame energy $E_p$ and  a lab frame photon momentum four-vector $M_p = (h\nu/c) (1,n^i)$ where $\nu$ is the photon frequency and {\bf n} the normalized propagation direction vector measured in the lab frame. The number of photons represented per packet is then $N = E_p/h\nu$.

In many cases, we initialize the radiation field based on the assumption of local thermodynamic equilibrium (LTE). This assumption is justified in most test problems we consider here, as the gas is optically thick to radiation across each zone. At the start of the calculation, a set number of packets, $N_{\rm init}$, are initiated in each zone. The radiation energy $a_r T_{0,g}^4 V_0$ is distributed equally among the packets in each zone. The packet comoving frequencies are sampled from a blackbody distribution at the local temperature and their directions are sampled isotropically in the comoving frame.

If necessary, it is possible to initialize photon packets without assuming LTE, as we will discuss for two test problems (section \ref{Sec:RadEqTest} and section \ref{Sec:Bondi}).

Photon packets are tracked in the lab frame, but the gas extinction coefficients and emissivities are calculated in the comoving frame. The extinction coefficient in the comoving frame $\chi_{0\nu}$ is then transformed into the lab frame using \citep{Thomas1930}
\begin{equation}
\chi_\nu =  \frac{\nu_0}{\nu} \chi_{0\nu} \, \, .
\label{Eq:ExtinctionTransform}
\end{equation}
While we have assumed that the comoving extinction is isotropic, in moving flows the
lab frame extinction $\chi_{\nu}$ is direction dependent (because of equations \ref{Eq:ExtinctionTransform} and \ref{Eq:EnergyDoppler}). 
The mean free path is longer for photons propagating along the flow and smaller
for photons propagating against it, a property that is 
essential to include to get the correct advection of radiation (see
section \ref{Sec:AdvectedPulse}).

The distance $l_k$ a photon travels in the lab frame before an interaction can be randomly sampled using
\begin{equation}
l_k = \chi_{\nu}^{-1} [-\ln(R)] \, \, ,
\end{equation}
where $R$ is a uniform random number between 0 and 1, not including 0. This distance
can be compared to the distance to the nearest cell boundary and the
distance to the end of the time-step ($\Delta t/c$) to determine the
next event. In an interaction event, a packet may be either scattered or absorbed, with
the probability of absorption at a given frequency denoted by $\epsilon_{\nu}$. More details about how photon interactions are implemented are given in section \ref{Sec:Interaction}

At each time-step, new packets may be generated to represent freshly radiated thermal energy. The emission of this energy provides the cooling contribution in equation \ref{Eq:ComovingG0}. The number of photon packets emitted in a zone over a lab frame time-step $\Delta t$ is
\begin{equation}
N_{\rm emit} = \frac{V \, \Delta t \, \epsilon \, \chi_{0P} \, c \, a_r \, T_{0,g}^4}{E_{0,p}} \, \, .
\label{Eq:PacketsEmitted}
\end{equation}
This expression will be modified slightly when implicit MCRT techniques are employed (section~\ref{Sec:ImplicitMC}). Here $V$ is the zone volume measured in the lab frame. We have made use of the fact that $V dt = V_0 dt_0$ \citep{Mihalas1984}, so to lowest order we may write $V \Delta t = V_0 \Delta t_0$. $E_{0,p}$ is the energy (not energy density) of each packet in the comoving frame. The value of $E_{0,p}$ can be chosen arbitrarily and ultimately sets the total number of packets included in a calculation. Typically, we choose the packet energy to be a small fraction of the zone energy, $E_{0,p} = 10^{-4} E_0 V_0$, however we limit the number of packets emitted per zone per time-step to a manageable maximum value (see Table~\ref{Tab:NumericalParameters}). The emitted packet's direction is sampled uniformly from an isotropic distribution in the comoving frame. The frequency of the packet is sampled from a distribution weighted by the comoving thermal emissivity, $\chi_{0\nu} \epsilon_\nu B_{0\nu}$. The packet energy, frequency and direction are then transformed into the lab frame using equations \ref{Eq:EnergyDoppler} and \ref{Eq:dtrans}. 

Radiative heating could, in principle, be evaluated by tallying the number of photon packets absorbed in each zone over a time interval. In any given time-step, however, the number of packets actually absorbed may be very small, especially if the medium is scattering dominated ($\epsilon_{\nu} \ll 1$). Instead, we can construct estimators in each cell of the comoving radiation energy density and radiation flux (equations \ref{Eq:ComovingEnergy} and \ref{Eq:ComovingFlux}) by summing over all path lengths of packets moving through the zone \citep{Lucy1999-1}
\begin{eqnarray}
\label{Eq:ComovingEnergyEstimator}
E_0 &= \frac{1}{c V \Delta t} \sum_p E_{p}\left(\frac{\nu}{\nu_0} \right)^2 l_p \\
\label{Eq:ComovingFluxEstimator}
F^i_0 &= \frac{1}{c V \Delta t} \sum_i E_{p} \left( \frac{\nu}{\nu_0} \right)^2 l_p n_0^i \, \, ,
\end{eqnarray}
where $E_{p}$ is the lab frame energy of packet $p$, $l_p$ is the length of the path the packet travels through the zone (which may be composed of multiple redirections), and we are again substituting $V \Delta t$ for $V_0 \Delta t_0$.

The flux estimator relies on the cancellation of packets moving in opposite directions, so it may be poorly sampled in practical calculations. As noted in Section \ref{Sec:Equations}, when the radiation is diffusing we may use the divergence of the lab-frame radiation pressure tensor to compute $G^i$ (equation \ref{Eq:PressureDivergence}). In this case, $P^{ij}$ is computed with the estimator 
\begin{equation}
\label{Eq:ComovingPressureEstimator}
P^{ij} = \frac{1}{c V \Delta t} \sum_p E_{p} l_p n^i n^j \,\, .
\end{equation}

\subsection{Interaction Physics}
\label{Sec:Interaction}

One advantage of MC transport methods is that it is relatively straightforward to simulate complicated physical interactions, such as anisotropic scattering, line transport, or polarization. In this section, we describe the treatment of select matter/radiation interactions.

\subsubsection{Absorption and Coherent Scattering}
\label{sec:scatter}
 
In the simplest of interaction events, a packet may be either coherently scattered or absorbed, with the probability of absorption at a given frequency, $\epsilon_\nu$, determined by the nature of the absorption interaction. In an explicit MC calculation, absorbed packets are simply removed from the calculation. In implicit MC calculations, some absorbed packets are not removed but instead undergo ``effective scattering'', as
will be described in Section~\ref{Sec:ImplicitMC}.

To simulate an isotropic, coherent scattering event, a packet is first Lorentz transformed to the comoving frame of the scatterer using \ref{Eq:EnergyDoppler} and \ref{Eq:dtrans}. A new direction is then sampled isotropically in the comoving frame, and the inverse transformation is applied to return the lab frame. In this process, the lab frame energy of the photon becomes
\begin{equation}
E_{\rm out} = E_{\rm inc} 
\frac{1 - n^i_{\rm inc} v^i/c}
{1 - n^i_{\rm out} v^i/c} \, \, ,
\label{Eq:Escatter}
\end{equation}
where $n^i_{\rm inc}$ and $n^i_{\rm out}$ are the incoming and outgoing packet direction vectors in the lab frame. The packet frequency changes in a corresponding way. When averaged over many scattering events, Eq.~\ref{Eq:Escatter} accounts for the adiabatic losses of the radiation field. Advection is captured via the anisotropy of the lab frame extinction coefficient $\chi$ and the outgoing direction vector $n^i_{\rm out}$. 

If desired, one can also take into account the random motions of scatterers, which may introduce additional Doppler shifts. In this case, the velocity vector of the individual scatterer must be randomly sampled at each interaction event. For example, the speed of a scatter could be randomly sampled from a Maxwell Boltzmann distribution with velocity dispersion $v_{\rm d} = (2 K T/m_s)^{1/2}$, where $m_s$ is the mass of the scatterer. The direction of the scatterer velocity vector is sampled from an isotropic distribution. The photon packet is then Lorentz transformed into the rest frame of the scatterer, a new propagation direction is chosen, and then the packet transformed back into the lab frame.

\subsubsection{Line Interactions}

The frequency-dependent cross-section of a line with rest frequency $\nu_0$ and oscillator strength $f_{\rm osc}$ is
\begin{equation}
\sigma(x) =  \frac{\sqrt{\pi} e^2}{m_e c} \frac{f_{\rm osc} }{\Delta \nu_{\rm d} } H(a,x) \, \, ,
\end{equation}
where $x$ is the frequency relative to line center in units of Doppler widths $x = (\nu - \nu_0)/\nu_{\rm d}$, where $\Delta \nu_{\rm d} = \nu_0 (v_{\rm d}/c)$ and we take the velocity dispersion $v_{\rm d}$ to be due to thermal line broadening. The line profile is described by the Voigt function
\begin{equation}
H(a,x) = \frac{a}{\pi} \int_{-\infty}^\infty   \frac{e^{-y^2}}{(x-y)^2 + a^2} dy \, \, .
\end{equation}
The parameter $a$ describes the importance of the wings relative to the core of the
line profile, and is a function of temperature
\begin{equation}
\label{Eq:VoigtParameter}
a = 4.7 \times 10^{-3}~ (T/10^4~{\rm K})^{-1/2} \, \, .
\end{equation}
We use the analytic fits for the Voigt profile provided by \citet{Tasitsiomi2006}.

The line absorption coefficient, $\alpha = n_l \sigma(x)$, depends on the the number density, $n_l$, of ions occupying the lower level of the transition, and hence requires knowledge of the ionization and excitation state of the gas. In the case of LTE, the state of the gas is readily determined by solving the Saha/Boltzmann equations. When LTE does not hold, the level populations must be determined by solving a  set of coupled rate equations, with the radiative transition rates estimated from the Monte Carlo transport. We postpone a discussion of the non-LTE problem, and assume here that the level population $n_l$ is known. 

To account for the thermal motions of ions, we randomly sample the ion velocities in a manner similar to that described in Section~\ref{sec:scatter}. However, since the line cross-section depends sensitively on frequency, the line-of-sight velocity $v_\parallel$ must be sampled from the modified distribution \citep{Zheng2002}
\begin{equation}
f(u_\parallel) = \frac{a}{\pi} \frac{e^{-u_\parallel^2}}{(x - u_\parallel)^2 + a^2} H^{-1}(a,x) \, \, ,
\end{equation}
where $u_\parallel = v_\parallel / v_d$. The transverse velocity components are sampled from the ordinary Maxwell-Boltzmann distribution
\begin{equation}
\begin{split}
v_{\perp,1} = v_{\rm d} \sqrt{-\ln(R_1)}\cos(2 \pi R_2) \, \, , \\
v_{\perp,2} = v_{\rm d} \sqrt{-\ln(R_1)}\sin(2 \pi R_2) \, \, ,
\end{split}
\end{equation}
where $R_1$ and $R_2$ are independently generated uninform random variables between 0 and 1, not including 0. $R_1$ sets the magnitude of the transverse velocity of the ion, and $R_2$ sets its direction in the transverse velocity plane. Packets can either be absorbed or scattered in a line, in the way described in Section~\ref{sec:scatter}. If desired, a treatment of fluorescence can also be included by randomly sampling the branching probability of de-excitation into all possible line transitions \citep{Lucy2002}. We will not discuss such a treatment here.

\section{Hydrodynamics}
\label{Sec:Hydro}

Our primary method for solving the hydrodynamics equations is a second-order Godunov scheme based on the PPM solver of \citet{Colella1984}. While this paper only presents results based on the one-dimensional version of this solver, we intend to extend it to higher spatial dimensions in a spatially unsplit manner following the description in \citet{Colella1990}. 

For 1-D spherically symmetric problems, we use a Lagrangian hydrodynamics solver because it allows for adaptive grid resolution. This solver is based on the von Neumann-Richtmyer staggered mesh scheme as described in \citet{Castor2004}.

As is well known, the inclusion of artificial viscosity is useful to damp numerical oscillations behind strong shocks, but has the negative effect of smearing out the shock front over a number of zones determined by a constant $C_q$, an adjustable parameter (see Table~\ref{Tab:NumericalParameters}). For the Lagrangian solver we can include a standard artificial viscosity of the form 
\begin{equation}
q = C_q \rho~{\rm max}( v_{\rm down} - v_{\rm up},0)^2 \, \, ,
\end{equation}
where $v_{\rm down}$ and $v_{\rm up}$ are the lower and upper zone velocities of the Lagrangian mass element. For the Godunov solver, the artificial viscosity is quite similar, although it involves modifying the numerical fluxes in the manner described in \citet{Lapidus1967}. We find that including artificial viscosity is helpful both in our Lagrangian solver and our Godunov solver, although for problems with strong shocks we can typically obtain similar results with a smaller value of $C_q$ in the Godunov case than in the Lagrangian case.

We use an operator-splitting procedure for coupling the radiation source terms to the hydrodynamics equations in a way similar to that of \citet{Noebauer2012}. Every time-step, we first perform the packet propagation through the fluid to construct the comoving frame estimators defined in equations \ref{Eq:ComovingEnergyEstimator} and \ref{Eq:ComovingFluxEstimator}. For the Eulerian version of the code, we use these estimators to construct the components of ${\bf G}$ in the lab frame, as given by equations \ref{Eq:G0Simplified} and \ref{Eq:GiSimplified}. We next use our Godunov solver to calculate the updates to the hydrodynamical state variables that would have occurred in the absence of radiation. Finally, we use our computed components of ${\bf G}$ to evaluate the right-hand sides of equations \ref{Eq:MomentumConservation} and \ref{Eq:EnergyConservation}. These source terms indicate the \emph{rate} at which momentum and energy are transferred per time per zone, so we multiply these rates by $V dt$ to compute the radiation contribution to energy and momentum for that time-step. In other words, our treatment of the radiative source terms is first order in time.

In the Lagrangian version of the code, we use a similar first-order approach to include the radiative source terms, but in this case we use the comoving quantity $cG_0$ for the rate of radiative heating or cooling. We multiply this rate by $V dt$ and add the result to the internal energy of the gas in each zone. For the radiative momentum contribution we use use the radial component of the radiative force. We multiply this force by $V dt$ and add this contribution to the total force that accelerates each zone boundary\footnote{This may introduce an error for moderately relativistic flows, because the velocities of the zone boundaries are measured in the lab frame, yet we are using a radiative force computed in the comoving frame to modify their acceleration}.

Future work might include the developments described in \citet{Miniati2007}, \citet{Sekora2010}, and \citet{Jiang2012} for including stiff radiative source terms more consistently within the Godunov solver. Indeed, once we have used MCRT to construct the radiation energy density, flux, and Eddington tensor, then the manner in which this information is incorporated into the Godunov solver could proceed in a manner identical that described in \citet{Jiang2012}. 

\section{Implicit Monte Carlo}
\label{Sec:ImplicitMC}

On the fluid flow time scale, the Monte Carlo simulation always provides a stable and accurate representation of the radiation field regardless of the time-stepping. However, an explicit treatment of the matter-radiation coupling will be unstable unless the time-steps are smaller than the time scale for radiative heating/cooling to significantly change the gas energy, given by 
\begin{equation}
t_{\rm rad} \approx \frac{1}{c \chi_{0P}} \frac{(\rho/\mu) k T_{0,g}/(\gamma_{\rm ad} - 1)}{a_r T_{0,g}^4} \, \, .
\label{Eq:CoolingTime}
\end{equation}
Under certain conditions, in particular cases where the radiation energy exceeds that of matter, $t_{\rm rad}$ may be much smaller than the Courant time-step. To avoid excessively small time-steps while maintaining stability, we implement the implicit Monte Carlo (IMC) methods first developed by \citet{Fleck1971} (see also \citet{Abdikamalov2012}). In this case one defines the Fleck factor
\begin{eqnarray}
f &\equiv& \frac{1}{1 + 4 \alpha_f \left(\frac{E_0}{e_0} \right) \left(c \,\Delta t \, \chi_{0P} \right)} \, \, ,
\end{eqnarray}
where $\alpha_f$ is a nondimensional parameter that can be given a value between 0.5 and 1 in order to ensure stability. The second term in the denominator can be thought of intuitively as the ratio of $\Delta t$ to $t_{\rm rad}$ (up to order unity factors).

The Fleck factor has several important roles. First, it is used to define an ``effective scattering'' rate. The true absorption fraction $\epsilon_{\nu}$ is multiplied by $f$ to determine a new probability that a photon interaction is treated by the code as an absorption event, rather than a scattering event. When $f \ll 1$ (i.e. the hydro time-step is much larger than $t_{\rm rad}$), then nearly all photon interactions are ffective scatttering interactions. Conversely, if $f \approx 1$ then the Fleck factor factor has little effect on the course of the simulation, and the probability of an absorption event remains approximately equal to $\epsilon_{\nu}$.

Second, the amount of thermal energy radiated each time step is also multiplied by $f$. This affects the second term in our calculation of $G^0_0$ in equation \ref{Eq:ComovingG0}, and the number of packets emitted each time step as given by equation \ref{Eq:PacketsEmitted}.

Finally, the Fleck factor is used to modify the process of adiabatic heating and cooling of the gas. Following \citet{Fleck1971}, we will define an adiabatic heating term $S^{\gamma}$:
\begin{equation}
S^{\gamma} = -(p + C_q) \frac{D}{Dt}\left(\frac{1}{\rho} \right) = (p + C_q) \frac{1}{\rho^2} \left(\frac{\partial \rho}{ \partial t} + v \frac{\partial \rho}{\partial x} \right) \, \, .
\end{equation}
This expresses the rate at which gas kinetic energy is converted into internal energy (or vice versa), and accounts for artificial viscosity. If desired, other heating source terms, such as energy released from nuclear reactions, could be added here. This rate $S^{\gamma}$ will also be multiplied by $f$, which amounts to subtracting $(1 -f)S^{\gamma}$ from the gas heating rate. In order to conserve energy, this same amount of energy per time step must then be added to the radiation field (which could amount to a negative contribution if $S^{\gamma}$ is negative).

In detail, for the Eulerian version of the code, we use the results of our Godunov scheme to construct the quantities $\partial \rho / \partial t$ and $\partial \rho / \partial x$, which are in turn used to consruct $S^{\gamma}$. We subtract $(1 -f)S^{\gamma}\rho \Delta t$ from the total energy density update of the hydro state vector\footnote{To be even more accurate we should transform the adiabatic heating/cooling rate, which is defined in the comoving frame, into a lab frame rate when performing this subtraction. Such a transformaion was not performed in this version of the code}. During the subsequent radiative transfer step, we add this contribution to the emission terms in equations \ref{Eq:ComovingG0} and \ref{Eq:PacketsEmitted}.

In the Lagrangian version of the code, when it is time to update the internal energy density of the fluid, we use $f S^{\gamma}$ for the amount of adiabatic heating or cooling, rather than the full $S^{\gamma}$ that we would use in the absence of implict Monte Carlo. The emission in the next radiative transfer step is modified in the same manner as in the Eulerian case.

\section{Radiation Test Problems}
\label{Sec:RadiationTests}

We have carried out a number of tests calculations to verify our code in a variety of physical situations.

\subsection{Frequency-dependent absorption with scattering}

First, we compare our MCRT implementation against analytic treatments of plane-parallel, semi-infinite, stratified, static atmospheres with frequency-dependent photon opacities. We follow the traditional convention of setting $\tau = 0$ at the observer's location at infinity, so that $\tau$ increases deeper into the atmosphere, along the $z$-axis of our coordinate system. Since the gas has zero bulk velocity in this test, we make no distinction between lab frame and comoving frame quantities for the rest of this subsection.

If, in addition to the assumptions listed in the previous paragraph, the source function is istropic, scattering is absent, and the temperature profile of the atmosphere is known, then the first three moments of the radiation intensity can be found exactly in terms of exponential integrals \citep[e.g.][]{Chandrasekhar1950, Kourganoff1952}. Here we follow \citet{Rutten2003} in writing the expressions for these moments as 
\begin{eqnarray}
E_{\nu}(\tau_{\nu}) &=& 2 \pi \int_0^{\infty} S_{\nu}(t_{\nu}) E_1 \left( |t_{\nu} - \tau_\nu | \right) dt_{\nu} \\
\label{Eq:E2Flux}
F_{\nu}^{z}(\tau_{\nu}) &=& 2 \pi \int_{\tau_{\nu}}^{\infty} S_{\nu}(t_{\nu}) E_2 \left( t_{\nu} - \tau_\nu \right) dt_{\nu} \nonumber \\
\label{Eq:F2Flux}
 &-& 2 \pi \int_0^{\tau_{\nu}} S_{\nu}(t_{\nu}) E_2 \left(\tau_\nu - t_{\nu} \right) dt_{\nu} \\ 
P_{\nu}^{zz}(\tau_\nu) &=& 2 \pi \int_0^{\infty} S_{\nu}(t_{\nu}) E_3 \left( |t_{\nu} - \tau_\nu \right|) dt_{\nu} \, \, ,
\end{eqnarray}
where
\begin{equation}
\label{Eq:ExponentialIntegral}
E_n(x) \equiv \int_0 ^1 e^{-x/\mu} \mu^{n - 1} \frac {d\mu}{\mu} \, \, .
\end{equation}

We next consider including a frequency-independent scattering extinction $\chi^s$, meant to represent electron scattering, in addition to the frequency-dependent absorption coefficient $\chi^t_{\nu}$. When scattering is included, an exact solution for the moments of the radiation intensity is rarely possible, although excellent approximate solutions can be derived, as we will now show.

It is conventional to introduce $J_{\nu} \equiv (c/4 \pi) E_{\nu}$, $H_{\nu} \equiv (1/4 \pi) F_{\nu}$ and $K_{\nu} \equiv (c/4 \pi) P_{\nu}^{zz}$. If all radiative cross sections are assumed to be isotropic, then the lowest two moment equations of the steady-state, plane-parallel transfer equation can be written
\begin{eqnarray}
\frac{dH_{\nu}}{d \tau_{\nu}} & = & \epsilon_{\nu}\left(J_{\nu} - B_{\nu} \right) \\
\frac{dK_{\nu}}{d \tau_{\nu}} & = & H_{\nu} \, \, .
\end{eqnarray}

If we employ the Eddington approximation, $K_{\nu} = J_{\nu}/3$, then the last two equations may be combined to yield \citep[e.g.][]{Rybicki1986}
\begin{equation}
\frac{d^2 J_{\nu}}{d \tau_{\nu}^2} = 3 \epsilon_{\nu} \left(J_{\nu} - B_{\nu} \right) \, \, .
\end{equation}
This is a linear, inhomogeneous ordinary differential equation for $J_{\nu}$. As such, it may be solved via the method of variation of parameters. \citet{Illarionov1972} present such a solution for the case when $\chi_{\nu}^t$ corresponds to \textit{bremsstrahlung}, so that it depends on both the density and temperature of the atmosphere at each depth. Here, we consider a slightly simpler situation in which $\epsilon_{\nu}$ is independent of depth. To specify the boundary conditions, we assume that $B_{\nu}$ approaches some finite value $B_{\nu,\infty}$ as $\tau_{\nu} \to \infty$, and that 
\begin{equation}
J_{\nu} = a_{\rm out} H_{\nu} \qquad {\rm at} \, \, \tau_{\nu} = 0 \, \, ,
\end{equation} 
where $a_{\rm out}$ is some constant value used to normalize the outgoing flux. If the two-stream approximation were to hold exactly as $\tau_{\nu} \to 0$, then $a_{\rm out}$ would equal $\sqrt{3}$ \citep{Rybicki1986}. 

In that case, the solution for $J_{\nu}$ becomes\footnote{Unlike \citet{Illarionov1972}, we allow the radiation to escape to $\tau_{\nu} = 0$, rather than cutting off the solution at $\tau_{\nu} = 1$. Additionally, we have not made the approximation $\chi^s_{0 \nu} \gg \chi^{t}_{0 \nu}$. }
\begin{eqnarray}
J_{\nu}(\tau_{\nu}) = e^{\tau_{\nu}\sqrt{3 \epsilon_{\nu}}}\int_{\tau_{\nu}}^{\infty} \frac{\sqrt{3 \epsilon_{\nu}}}{2} \, B_{\nu}(t_{\nu}) e^{-t_{\nu} \sqrt{3 \epsilon_{\nu}}}dt_{\nu} \nonumber \\ + e^{-\tau_{\nu}\sqrt{3 \epsilon_{\nu}}} \Bigg [ \int_0^{\tau_{\nu}} \frac{\sqrt{3 \epsilon_{\nu}}}{2} \, B_{\nu}(t_{\nu}) e^{t_{\nu}\sqrt{3 \epsilon_{\nu}}}dt_{\nu} \nonumber \\ - \frac{1 - a_{\rm out} \sqrt{\frac{\epsilon_{\nu}}{3}}}{1 + a_{\rm out} \sqrt{\frac{\epsilon_{\nu}}{3}}}  \int_{0}^{\infty} \frac{\sqrt{3 \epsilon_{\nu}}}{2} \, B_{\nu}(t_{\nu}) e^{-t_{\nu} \sqrt{3 \epsilon_{\nu}}}dt_{\nu} \Bigg] \, \, .
\end{eqnarray}

The emergent flux can then be computed as
\begin{eqnarray}
\label{Eq:FluxEddington}
F_{\nu}(0) &=& 4 \pi H_{\nu}(0) = \frac{4 \pi}{3} \frac{d J_{\nu}}{d \tau_{\nu}}(0) \nonumber \\
&=& 4 \pi \frac{\epsilon{_\nu}}{1 + a_{\rm out}\sqrt{\frac{\epsilon_{\nu}}{3}}}\int_{0}^{\infty}B_{\nu}(t_{\nu}) e^{-t_{\nu} \sqrt{3 \epsilon_\nu}} dt_{\nu} \, \, .
\end{eqnarray}

Since the Eddington approximation was used to derive equation \ref{Eq:FluxEddington}, taking its limit as $\epsilon_{\nu} \to 1$ (i.e., in the limit of no scattering) does not recover equation \ref{Eq:F2Flux} evaluated at $\tau_{\nu} = 0$. However, if we consider equation \ref{Eq:F2Flux} under the the two-stream approximation so that $\mu$ is fixed at $1/\sqrt{3}$ in equation \ref{Eq:ExponentialIntegral}, and $a_{\rm out} = \sqrt{3}$, then we do indeed recover the $\epsilon_{\nu} \to 1$ limit of equation \ref{Eq:FluxEddington}.

To gain insight into equation \ref{Eq:FluxEddington}, we consider the thermalization depth
\begin{equation}
\label{Eq:ThermalizationDepth}
\Lambda_{\nu} \equiv 1/ \sqrt{\epsilon_{\nu}} \, \, .
\end{equation}
This is the average depth that a freshly emitted photon with frequency $\nu$ will travel via scattering before being reabsorbed. The heuristic derivation (following \citet{Rutten2003}) for equation \ref{Eq:ThermalizationDepth} is as follows. During each scattering event, the probability that the photon is absorbed is $\epsilon_{\nu}$, by definition. Thus, an emitted photon will scatter an average of $1/\epsilon_{\nu}$ times before being reabsorbed. Meanwhile, for any random-walk process, the mean displacement of a packet that has undergone $N$ re-directions, each of mean free path $l$, is approximately $l \sqrt{N}$. Consequently, the average distance between emission and absorption events is $l/ \sqrt {\epsilon}$. Converting this distance to an optical depth gives us our result. A factor of $\sqrt{3}$ in front of $\epsilon_{\nu}$ can account for the average angle with respect to the $z$-axis along which the photons travel in the Eddington approximation.

We do not expect escaping photons to have been emitted at temperatures corresponding to optical depth much greater than the thermalization depth. In other words, frequencies with large thermalization depths allow us to see such photons that were emitted from deeper, hotter portions of the atmosphere. 

We ran three test calculations, each with a different degree of scattering, to test the code against these solutions. In all three cases, we used a domain of total height $h~=~10^{14}$~cm. For the first two tests we divided the domain into 128 zones of equal height, and for the final test we used 256 zones. The domain was filled with gas following a power-law density profile
\begin{equation}
\rho(z) = \rho_{\rm max} \left[1 + \left(\frac{h - z}{z_s} \right)^{p}\right]^{-1} \, \, ,
\end{equation}
where we have chosen $\rho_{\max}~=~2.09~\times~10^{-11}$~g~cm$^{-3}$ (yielding an optical depth to electron scattering of 100, where the electron scattering opacity is 0.4 cm$^2$ g$^{-1}$ for fully ionized hydrogen), $z_s~=~10^{13}$~cm, and $p~=~3$. Photons were emitted from the $z~=~h$ plane and propagate toward the $z~=~0$ plane, where they are tallied to generate an outgoing spectral energy distribution (SED). Any photons that scattered back past the $z~=~h$ plane were treated as absorbed by the luminous source and were no longer tracked. We adjusted the photon flux from the inner emitting surface (at $z = h$) so that the bolometric, steady-state radiative flux escaping to infinity would equal a constant value of $1.64 \times 10^{20}$ erg s$^{-1}$ cm$^{-2}$ in all three calculations.

We chose a normalization for $\chi^t_{\nu}$ so that it would match the electron scattering extinction at 100 Angstroms. We also let $\chi^t_{\nu}$ scale as $\nu^{-1}$. Our wavelength resolution was set by dividing the interval between 1 and $10^4$ Angstroms into 100 bins equally spaced logarithmically. 

Anticipating that the densities and temperatures in these calculations would correspond to cooling times that were orders of magnitude shorter than the radiative diffusion time through the computational domain, we used a fully implicit treatment of the radiative heating and cooling. Absorption events were always treated as effectively scattered, and we periodically re-computed the temperature of the gas in each zone by enforcing radiative equilibrium until a steady state was reached. 

\begin{figure*}
\includegraphics[width=\textwidth]{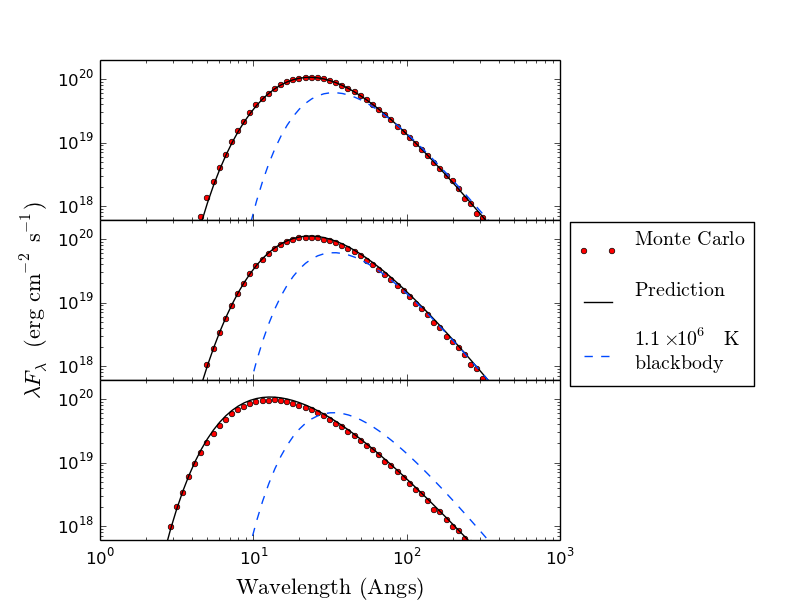}
\caption{Tests of outgoing SEDs for static, stratified, plane-parallel atmospheres with frequency-dependent opacities. Top panel: No scattering. The absorption coefficient $\chi_{\nu}^t$ is chosen so that it matches the Thomson extinction at $100$ Angstroms, and declines as $\nu^{-1}$. The exact analytic solution used for comparison is given in equation \ref{Eq:F2Flux}. The blackbody spectrum is included to guide the eye and to illustrate how the emergent flux in this calculation includes emission from gas layers at a range of temeperatures. Middle panel: Thomson scattering has been added as a contribution to the opacity, but all other details of the calculation remain the same as the top panel. The analytic prediction now uses the Eddington approximation and is given by equation \ref{Eq:FluxEddington}. Bottom panel: The thermal opacity is now multiplied by a factor of 0.01, but all other details remain the same as in the middle panel. For sufficiently small $\epsilon_{\nu}$, as in this panel, the SED shifts toward smaller wavelengths even while the peak value of $\lambda F_{\lambda}$ remains nearly the same as in higher $\epsilon_{\nu}$ runs. The slight over-prediction of flux in this case seems to improve as spatial resolution is increased. Higher spatial resolution is needed in this case because the photons that escape were initially emitted from deeper, hotter portions of the atmosphere with higher temperature gradients than in the previous two cases. }
\label{Fig:SedAbsorptionAndScattering}
\end{figure*}

Figure \ref{Fig:SedAbsorptionAndScattering} shows the outgoing SED for three test cases. In the first case, shown in the top panel, we used only the absorption coefficient $\chi^t_{\nu}$, and neglected scattering entirely. This allowed us to solve for the outgoing flux by invoking equation \ref{Eq:E2Flux} at $\tau_{\nu} = 0$ for all $\nu$. The match between the analytic formula and the Monte Carlo results is excellent.

\begin{figure}
\includegraphics[width=0.5\textwidth]{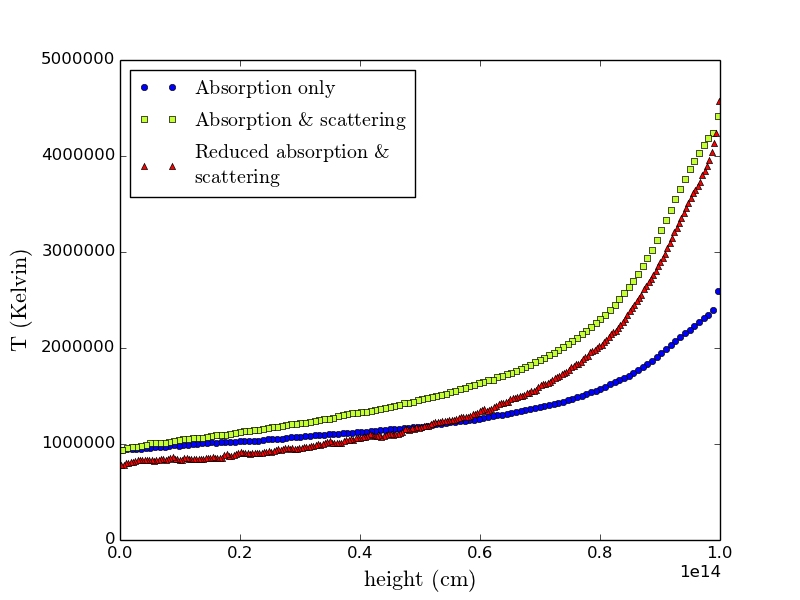}
\caption{The gas temperature as a function of height for the static atmosphere tests. The three curves correspond to the three panels in Figure \ref{Fig:SedAbsorptionAndScattering}. Although the temperatures are noticeably different between all three runs, the outgoing flux has been adjusted to be the same in all three cases.}
\label{Fig:StaticAtmosphereTempVsRadius}
\end{figure}

In the second case, shown in the middle panel of Figure \ref{Fig:SedAbsorptionAndScattering}, we add Thomson scattering but we keep all other details of the simulation the same as before. Given our functional form for $\chi^t_{\nu}$ described above, $\epsilon_{\nu}$ ranges from $10^{-2}$ at 1 Angstrom to 0.9 at $1000$ Angstroms. As shown in Figure \ref{Fig:StaticAtmosphereTempVsRadius}, the inclusion of scattering along with absorption, while forcing the escaping flux to be the same, leads to higher temperatures in all regions of the atmosphere. Remarkably, this temperature adjustment occurs in such a way as to keep the shape of the outgoing SED nearly identical to the case without scattering (compare the first and second panels of Figure \ref{Fig:SedAbsorptionAndScattering}). The Eddington approximation prediction for the shape of the SED (equation \ref{Eq:FluxEddington}) still matches the computed SED quite well.

Finally, the bottom panel of Figure \ref{Fig:SedAbsorptionAndScattering} shows the results of another test that includes both absorption and scattering, but this time the absorption opacity is reduced to a value of 0.01 times the value we had used previously. Now $\epsilon_{\nu}$ ranges from $10^{-4}$ at 1 Angstrom to 0.09 at $1000$ Angstroms. In this case there is a slight drop in temperature compared to the previous case at all depths in the atmosphere, as seen in Figure \ref{Fig:StaticAtmosphereTempVsRadius}. However, this time the SED shifts markedly in the blueward direction, which is evident in the bottom panel of figure \ref{Fig:SedAbsorptionAndScattering}. This can be understood in terms of the thermalization length described earlier. As the atmosphere becomes increasingly scattering dominated, the photons that escape to the observer tend to have been emitted at higher Thomson optical depth, where the temperature is higher. Again, agreement with the analytic formula is very good, verifying the MCRT calculation of multi-frequency transport in a scattering dominated regime.

\subsection{Line Transport}

\begin{figure}
\includegraphics[width=0.5\textwidth]{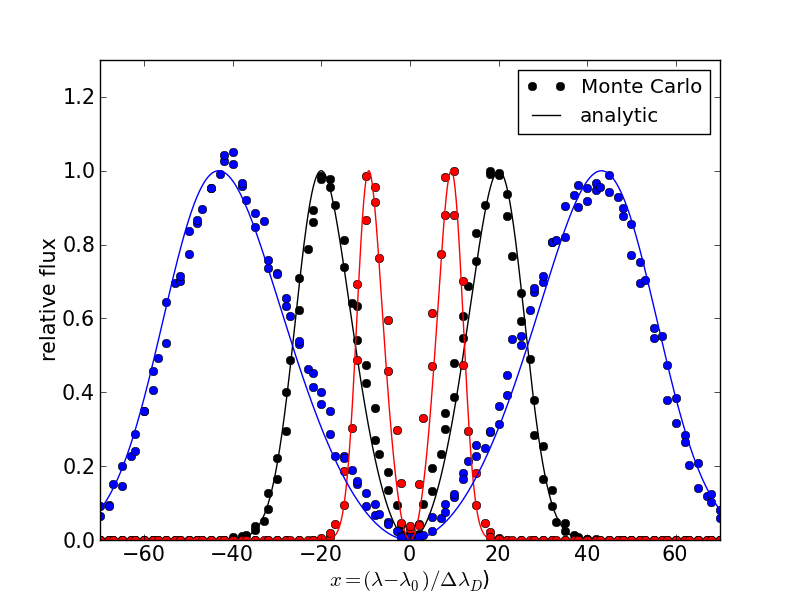}
\caption{
\label{Fig:static_line}
Test of line transport in a static medium, comparing Monte Carlo results (circles) to an analytic solution based on the diffusion approximation (Equation~\ref{eq:static_line}, solid lines). In this problem, a point source radiates line photons into a uniform spherical medium with a pure-scattering optical depth at line center of  $\tau_c = 10^4$ (red), $\tau_c = 10^5$ (black) and $\tau_c = 10^6$ (blue).
}
\end{figure}

\begin{figure}
\includegraphics[width=0.5\textwidth]{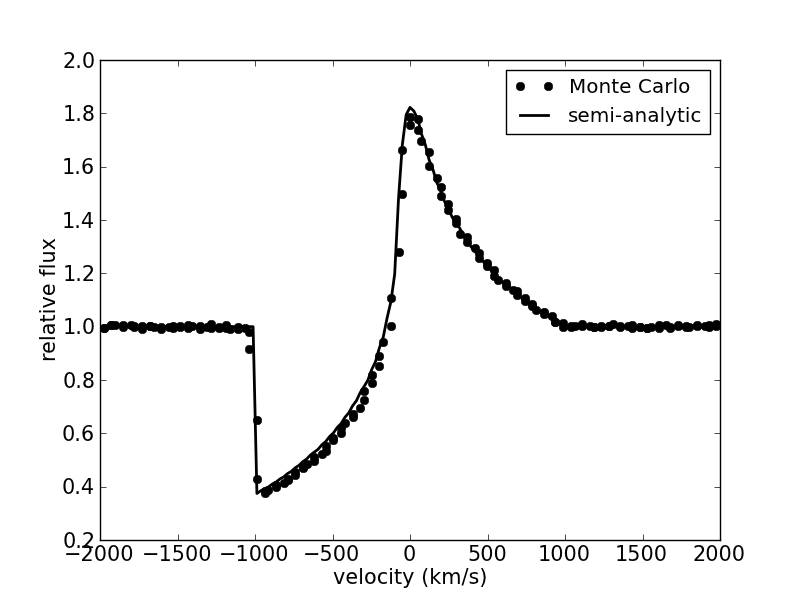}
\caption{Test of line transport in a moving homologously expanding medium. A spherical source radiates continuum photons into a uniform pure-scattering medium with Sobolev line optical depth of $\tau_s = 1$. Results from the Monte Carlo (circles) are compared to the semi-analytic solution based on the Sobolev approximation (Equation~\ref{eq:moving_line}).
\label{Fig:moving_line}}
\end{figure}

We next test the transport of line radiation in both moving and static media. We use the spherical Lagrangian version of the code and inject photons into a uniform density sphere  of radius $r_{\rm max} = 10^{15}$~cm. The thermal motions of ions are taken into account, with a velocity dispersion of $v_{\rm d} = 25~{\rm km~s^{-1}}$. The line opacity is taken to be pure-scattering, and for computational expediency we adopt a large Voigt parameter of $a = 0.1$ (see equation \ref{Eq:VoigtParameter}).

In the first test, we consider a static atmosphere with a total radial optical depth at line center of $\tau_c$. Photons are injected at the center of the sphere and at the  line center rest frame frequency $\nu_c$. An analytic solution to the line scattering problem in the plane parallel case was derived by \citet{Neufeld1990} under the Eddington approximation, and generalized to a spherical atmosphere by \citet{Dijkstra2006}, who find a total flux density at the surface of the sphere
\begin{equation}
J(x) = \frac{\sqrt{\pi}}{24 \sqrt{\pi a \tau_c}} \biggl[ \frac{x^2}{1 + \cosh[\sqrt{2 \pi^3/27}(|x^3|/a\tau_c)]} \biggr]
\label{eq:static_line}
\end{equation}
In Figure~\ref{Fig:static_line} we show results of the MC transport for spheres of optical depth $\tau_c = 10^4$ and $10^6$. The resulting line features show a characteristic double peaked profile. This is because photons are Doppler shifted by the thermal motions of the scatterers, and preferentially escape in the less opaque line wings. Our MC results show favorable agreement with the analytic solution Equation~\ref{eq:static_line}, comparable to those seen in other MCRT line transport codes \citep{Dijkstra2006}.

To test line transport in a moving atmosphere, we consider the case where the sphere of gas is expanding homologously (i.e., velocity proportional to radius). We emit photon packets from the surface of a spherical inner boundary of uniform specific intensity $I_p$ in the lab frame at a radius $r_p = 10^{14}$~cm. The velocity structure is given by $v(r) = v_{\rm max} (r/r_{\rm max})$, with  $v_{\rm max} = 10^8 {\rm cm~s^{-1}}$. Because the velocity scale height of this problem is much greater than the ion thermal velocities, the Sobolev approximation applies. The emergent line profile in the lab frame is then given by an integral over the impact parameter $p$, \citep[e.g.,][]{Jeffery1990}
\begin{equation}
F(\nu) = 2 \pi \int_{0}^\infty 
\biggl[ I_p e^{-\tau_{\rm s}} + S(r) (1 - e^{-\tau_{\rm s}}) \biggr] pdp
\label{eq:moving_line}
\end{equation}
where the Sobolev optical depth is 
\begin{equation}
\tau_s(r) = \frac{\pi e^2}{m_e c}  \frac{c}{\nu_c} \frac{f_{\rm osc} n_l}{dv/dr}.
\end{equation}
In the present example the velocity gradient is $dv/dr = v_{\rm max}/r_{\rm max}$. The source function for a pure-scattering line is equal to the mean intensity of the
radiation field, $S(r) = J(r) = W(r) I_p$ where the dilution factor is
\begin{equation}
W(r) =  \frac{1}{2} \biggl[ 1 - \sqrt{1 - (r_p/r)^2} \biggl] \, \, .
\end{equation}
As discussed in \cite{Jeffery1990}, to properly treat the boundary condition of the photosphere, $\tau_s(r)$ and $S(r)$ are zero for the spatial region inside and behind the photosphere, while $I_p(p)$ is zero for $p > r_p$.

Figure~\ref{Fig:moving_line} shows results for a constant density atmosphere with  $\tau_{s} = 1$. The spectrum of the MCRT code, which resolves the line profile, is in good agreement with the Sobolev semi-analytic solution.

\section{Radiation-hydrodynamics Test Problems}
\label{Sec:RadhydroTests}

We next discuss test problems in which the energy and momentum coupling of the gas and radiation is considered. In what follows, we define the radiation temperature as $T_{0,r} = (E_0/a_r)^{1/4}$, where $E_0$ is the comoving radiation energy density. We use an ideal gas equation of state with $\gamma_{\rm ad} = 5/3$.

Table \ref{Tab:NumericalParameters} lists the numerical parameters used in each radiation-hydrodynamics test problem.

\begin{deluxetable*}{ccccccccc}
\tablecaption{Numerical parameters}
\tablehead{\colhead{Test} & \colhead{\# of zones} & \colhead{Zone width} & \colhead{$dt$} & $t_{\rm stop}$ & \colhead{$\alpha_f$} & \colhead{$C_q$} & \colhead{Initial packets} & \colhead{Max packets emitted} \\  \colhead{} & \colhead{} & \colhead{(cm)} & \colhead{} & \colhead{(s)} & \colhead{} & \colhead{} & \colhead{per zone} & \colhead{per zone per step}  }
\startdata
Radiative equilibrium & 2 & $ 5.0 \times 10^9$ & $1.0 \times 10^{-11}$ s & $1.0 \times 10^{-7}$ & 0. & 0. & 1 & 0  \\
(no IMC) &  &  &  &  &  &  &  &   \\ \\
Radiative equilibrium & 2 & $ 5.0 \times 10^9$ & $1.0 \times 10^{-14}$ s & $1.0 \times 10^{-7}$ & 0.5 or & 0. & 1 & 0  \\
(IMC) &  &  &  &  & 1.0 &  &  &   \\ \\ 
Advected pulse & 201 & 0.00995 & $1.0 \times 10^{-13}$ s & $1.0 \times 10^{-10}$ & N/A & 0. & $10^5$ & 0\\ 
 &  &  &  &  &  &  & (center zone only) &   \\ \\
Homologous expansion & 64 & Variable & CFL 0.2 & $1.0 \times 10^5$ & N/A & 0. & 10 & 0 \\ \\
Bondi accretion & 2048 & Variable & CFL 0.2 & $3.0 \times 10^6$ & 1.0 & 0. & 10 & 40 \\ 
 &  &  &  &  &  &  &  & (from source only)   \\ \\
${\cal M} = 2$ steady shock & 512 & $5.86 \times 10^{-5}$ & CFL 0.5 & $1.0 \times 10^{-9}$ & 1.0 & 0.1 & 10 & 400 \\ \\
${\cal M} = 5$ steady shock & 2048 & $1.95 \times 10^{-5}$ & CFL 0.5 & $1.9 \times 10^{-9}$ & 1.0 & 0.1 & 10 & 100 \\ \\
${\cal M} = 70$ steady shock & 896 & $1.29 \times 10^{-3}$ & CFL 0.1 & $1.0 \times 10^{-9}$ & 1.0 & 0.1 & 7680 & 7680 \\ \\ 
Sub-Critical & 512 & $1.37 \times 10^8$ & CFL 0.5 & $4.0 \times 10^4$ & 0. & 0.5 & 1000 & 4000 \\
moving shock (Ensman) &  &  &  &  &  &  &  & \\ \\
Super-Critical & 512  & $1.37 \times 10^8$ & CFL 0.2 & $1.3 \times 10^4$ & 1.0 & 0.5 & 1000 & 1000 \\
moving shock (Ensman) &  &  &  &  &  &  &  & \\ \\

\enddata
\tablecomments{Numerical parameters. When zone width is listed as ``variable'', the Lagrangian version of the code is being used. When $\alpha_f$ is listed ``N/A'', the radiation energy and gas are thermally decoupled (i.e. $\epsilon$ = 0 so that radiation always scatters and is never absorbed).}
\label{Tab:NumericalParameters}
\end{deluxetable*}

\subsection{Evolution to radiative equilibrium}
\label{Sec:RadEqTest}

We begin with a standard test of the heating and cooling of the gas by radiation, which also provides clear a demonstration of the application of implicit MC techniques. We chose here a setup identical to that of \citet{Turner2001}, although modified versions of the test have appeared elsewhere, including \citet{Noebauer2012}.

In this test we use the Eulerian version of the hydro solver. We again consider gas with zero bulk velocity, so that the lab frame and the comoving frame are identical, although we retain the comoving frame notation. The computational domain is filled with static gas at a uniform density of $\rho_0 = 10^{-7}$ g cm$^{-3}$, a mean atomic mass of $\mu = 0.6$, and a gray opacity $\kappa_0 = 0.4$ cm$^{2}$ g$^{-1}$. Additionally, a uniform and isotropic radiation field is initialized with energy density $10^{12}$ erg cm$^{-3}$, so that  $T_{0,r} = 3.4 \times 10^6$ K. Here $\epsilon = 1$, so that the gas and radiation are fully thermally coupled. Although the radiation pressure overwhelms the gas pressure in this test, the radiation field is isotropic, so the radiation pressure does not accelerate the gas. Reflecting boundary conditions were used for the radiation.

In this context, the gas energy equation (Equation~\ref{Eq:EnergyConservation}) simplifies to 
\begin{eqnarray}
\frac{d e_0}{dt} &= \chi c a_r T_{0,r}^4 - 4 \chi B(T_{0,g}) \nonumber \\ 
&= \chi c a_r(T_{0,r}^4 - T_{0,g}^4)
\label{eq:radeq}
\end{eqnarray}
where $B$ is the frequency-integrated Planck function \citet{Turner2001}. We consider two versions of the test, one in which the gas is heated by radiation, and another in which the gas cools. For the heating case, the gas is given an initial thermal energy density of $10^2$ erg cm$^{-3}$, corresponding to $T_{0,g} = 11$ K. For the cooling case, the initial thermal energy density is $10^{10}$ erg cm$^{-3}$, corresponding to $T_{0,g} = 1.1 \times 10^9$ Kelvin. In both  cases, the radiation energy greatly exceeds the gas energy density, and so remains
nearly constant during the energy exchange. This means that the gas will ultimately heat or cool to reach the radiation temperature, corresponding to an equilibrium gas energy density of $7.8 \times 10^7$ erg cm$^{-3}$.

This test only follows the evolution of the gas up to an elapsed time of $10^{-5}$ s, whereas the photon interaction time $1/(\rho \kappa c)$ is approximately $10^{-3}$ s. In the absence of explicit photon interactions, the energy exchange between gas and radiation is deterministic, and so the number of photon packets employed has no effect on the solution.

\begin{figure*}
\includegraphics[width=0.75\textwidth]{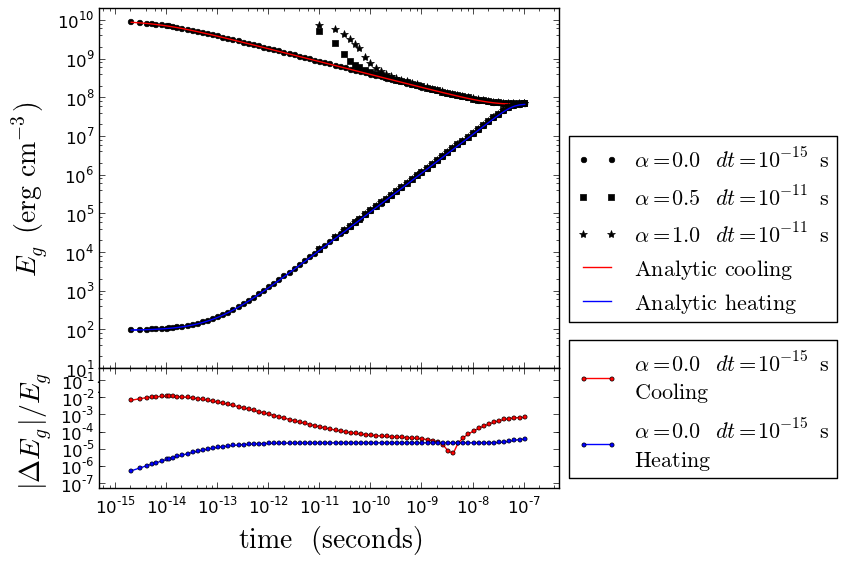}
\caption{Tests of the approach to radiative equilibrium in a radiation-energy dominated gas, with gray radiative opacity. The red and blue curves represent the analytic gas heating and cooling curves as computed from equation \ref{eq:radeq}. The points represent values computed from the Monte Carlo simulation for three sets of numerical parameters as described in the legend. All other numerical parameters are held at the values specified in Table~\ref{Tab:NumericalParameters}. Two different implicit treatments of the heating and cooling are used for large time-steps, in addition to an explicit numerical treatment at time-step much shorter than the cooling time. The bottom panel shows the fractional error compared to the analytic solution for the case of explicit heating and cooling.}
\label{Fig:RadEq}
\end{figure*}

Figure~\ref{Fig:RadEq} displays the gas heating and cooling curves compared to the analytic solution of Equation~\ref{eq:radeq}. Consider the cooling curve first. According to equation \ref{Eq:CoolingTime}, the cooling time at the beginning of the simulation is $1.7 \times 10^{-15}$ s. If we take a time-step smaller than this, such as $10^{-15}$ seconds, then no implicit methods are needed, and the gas temperature follows the analytic cooling curve to an accuracy of better than 1.3\% at all times. If we wish to take much larger time-steps, then we must turn on the implicit Monte Carlo by setting $\alpha_f \ge 0.5$, otherwise the code generates negative temperatures and crashes after the first step of the calculation. Figure~\ref{Fig:RadEq} shows the results of taking $\alpha_f = 0.5$ and $\alpha_f = 1.0$ for $dt = 10^{-11}$ s. In both cases, the cooling curves approach the analytic solution after many time-steps, but the cooling is artificially slow at early times. The $\alpha_f = 0.5$ case converges to the correct solution more quickly than the $\alpha_f = 1.0$ case, demonstrating that one should strive for the smallest value of $\alpha_f$ that still maintains stability.

The heating curves follow the analytic solution to within one part in $10^{-4}$ at all times, regardless of the value of $\alpha_f$ chosen or the size of the time-step up to $10^{-11}$ that we tested, although larger time-steps could be used for the heating case.

\subsection{Advected radiation pulse}
\label{Sec:AdvectedPulse}

In a moving, optically thick medium, radiation should be swept along with the matter. This represents an important and non-trivial test of the MCRT routine, as advection is not explicitly included in the code. Instead, advection is a statistical consequence  of the lab frame anisotropy of the lab frame extinction coefficient and scattering function. When averaged over many scatters, these effects preferentially guide packets upstream. 
 
Our test is similar to the radiation diffusion tests presented in \citet{Harries2011} and \citet{Noebauer2012}, but with the added effect of advection. We use the Eulerian version of the code, and consider a homogeneous gas distribution  from $x = -1$ cm to $x = 1$~cm. The gas is pure-scattering ($\epsilon = 0$) with $\mu = 0.5$, and is given a uniform lab frame velocity of $2 \times 10^9$ cm s$^{-1}$. The scattering opacity is taken to be $\kappa_0 = 10^9$ cm$^2$ g$^{-1}$, which gives an optical depth across each zone equal to $1$ in the comoving frame. The radiation is initialized isotropically in the comoving frame of the central zone only, with a comoving energy density of $10^{10}$ erg cm$^{-3}$. 

\begin{figure}
\includegraphics[width=0.5\textwidth]{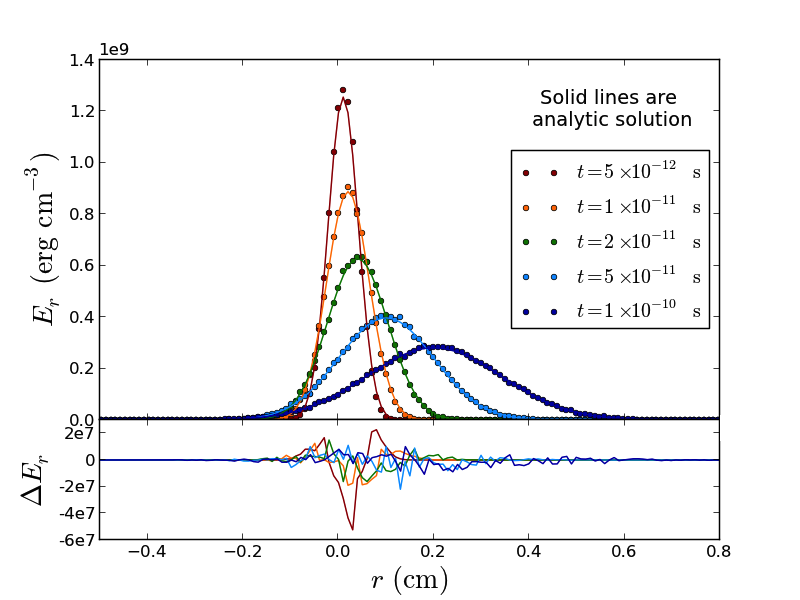}
\caption{Test of advection and diffusion of radiative energy in a moving fluid with gray scattering opacity. The mean free path of the photons is approximately $0.01$ cm, so the fluid is highly optically thick to the radiation and sweeps the radiative energy along with it. The analytic solution is given by the advection-diffusion equation, and the bottom panel shows the absolute error in the computed radiative energy density as compared to the analytic solution. The numerical parameters used in this test are specified in Table~\ref{Tab:NumericalParameters}.}
\label{Fig:AdvectingPulse}
\end{figure}

Since the gas and radiation are thermally decoupled in this test, and we are primarily interested in the advection and diffusion of the radiation energy, the value chosen for the gas temperature is arbitrary. However, a lower temperature results in a higher mach number. In the test corresponding to Figure \ref{Fig:AdvectingPulse} (discussed below), we chose to set the gas temperature to $10^4$, corresponding to an isothermal mach number of roughly 1560. In this case, we found it necessary to include a floor for the gas energy to prevent it from dropping below zero. Radiation is allowed to escape through either side of the domain, and periodic boundary conditions are employed for the hydrodynamics solver.

As discussed in \citet{Harries2011} and \citet{Noebauer2012}, the evolution of the radiation energy can be solved for analytically in the diffusion approximation. Figure~\ref{Fig:AdvectingPulse} compares our computed radiation temperature to this solution at various times. One sees that the radiation pulse moves along with the gas at the expected velocity. We confirmed that {\it both} the transformation of the extinction coefficient (Equation~\ref{Eq:ExtinctionTransform}) and the effect of aberration (Equation~\ref{Eq:dtrans}) must be included to reproduce the proper advection velocity. Thus, even in problems with velocities $v \ll c$, a special relativistic MC treatment is desirable to recover the proper advection behavior.

\subsection{Opaque Expanding sphere}

We next consider a problem designed to test whether the code properly handles radiation  energy losses due to expansion. This is also a non-trivial test of the MCRT routine, as no explicit term for radiation $pdV$ work is included in the code. Instead, the change in the radiation energy density is a statistical result of the multiple Doppler shifts photon packets incur as they  scatter anisotropically off of  moving gas. 

We consider a spherical gas cloud undergoing homologous expansion  (i.e., velocity proportional to radius) and opaque enough  that photons  do not diffuse significantly, but are rather advected along with the flow. Such an adiabatically expanding flow cools as $T \propto V^{1 - \gamma_{\rm ad}}$, with $V \propto r_{\rm out}^3$. We assume the medium is pure-scattering  ($\epsilon = 0$), so that the radiation and gas are thermally decoupled. Hence the gas ($\gamma_{\rm ad} = 5/3$) should evolve as $T_{0,g} \propto r_{\rm out}^{-2}$ while the radiation ($\gamma_{\rm ad} = 4/3$) should evolve separately as $T_{0,r} \propto r_{\rm out}^{-1}$. 

For this test, we use the Lagrangian version of the hydro solver. The outer edge of the computational domain expands homologously as $r_{\rm out} = r_{\rm out, i} + v_{\rm out} t$ , where $t$ is the time elapsed. We take $r_{\rm out, i} = 10^{13}$~cm and $v_{\rm out} = 10^{9}~{\rm cm~s}^{-1}$. The gas is initially uniform with a temperature of $10^4$~K, a density of $\rho = 4.75 \times 10^{-7}$ g cm$^{-3}$, $\mu = 0.5$, and $\kappa_0 = 0.4$ cm$^2$ g$^{-1}$. Reflecting boundary conditions at $r_{\rm out}$ are used for the radiation. To compute the fluid pressure gradient at the outer boundary we linearly extrpolate the pressure from the outermost two zones to evaluate the pressure beyond the outermost radial zone, although the gas pressure does not play an important role in this test.

\begin{figure}
\includegraphics[width=0.5\textwidth]{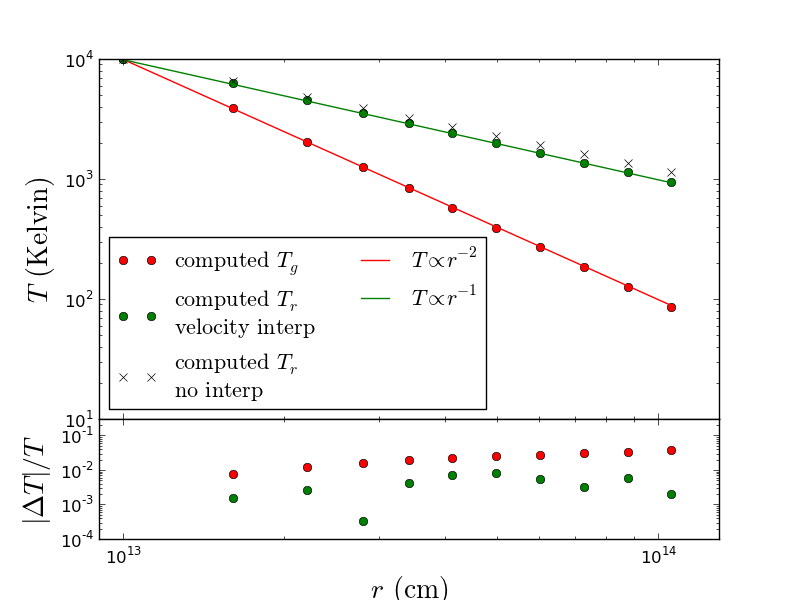}
\caption{Test of the temperature evolution of a homologously expanding sphere of fluid that is optically thick to scattering radiation. Since the gas and radiation have different adiabatic indices, their temperatures as a function of radius/time follow different relationships. We find that in order to achieve a percent-level match to the expected temperature profiles, we must interpolate the velocity of the fluid between neighboring Lagrangian mass cells. The fractional error in the computed versus expected temperatures is shown in the bottom panel. The numerical parameters used in this test are listed in Table~\ref{Tab:NumericalParameters}.}
\label{Fig:HomologousExpansion}
\end{figure}

Figure~\ref{Fig:HomologousExpansion} displays the spatially-averaged gas and radiation temperatures as a function of $r_{\rm out}$ (each zone was given equal weight in the average). The code recovers the expected adiabatic loses of the gas and radiation field. We ran two versions of this test, one in which the gas velocity was taken to be piece-wise constant in each zone, and the other in which the gas velocity was linearly interpolated within each zone. 

As is evident in the figure, the evolution of the radiation temperature is more accurately computed for the case in which velocity interpolation was used, indicating that an adequate resolution of the gas velocity field is necessary properly calculate the radiation $pdV$ work.

\subsection{Bondi accretion with optically thin radiation pressure}
\label{Sec:Bondi}

The classic Bondi problem of steady-state, spherically symmetric gravitational accretion \citep{Bondi1952} provides an opportunity for us to test the effect of radiation force in the optically thin limit. Our treatment of the problem closely follows that of \citet{Krumholz2007}. For an accreting object with mass $M$ and isotropic radiative luminosity $L$ we may define the Eddington factor  
\begin{equation}
f_{\rm Edd} = \frac{\kappa_0 L}{4 \pi G M c},
\end{equation}
 where $\kappa_0$ is the gas opacity, taken here to be gray. We consider the isothermal case. For a sound speed $c_s$, we may then define the radiatively-inhibited Bondi radius as 
\begin{equation}
r_{\rm B} = (1 - f_{\rm Edd})\frac{G M}{c_s^2}.
\end{equation}
The expected steady-state mass accretion rate is then
\begin{equation}
\dot{M}_{\rm B} = 4 \pi \left(\frac{e^{3/2}}{4}\right) c_s \rho_{\infty}r_{\rm B}^2,
\label{Eq:BondiRate}
\end{equation}
where $\rho_{\infty}$ is the gas density at the outer boundary of the domain.

We set $M = 10~M_\odot$, $L = 1.63 \times 10^5$ $L_{\odot}$, $c_s = 1.29 \times 10^7$ cm s$^{-1}$, $\rho_{\infty} = 10^{-18}$ g cm$^{-3}$,  $\mu = 1$, and $\kappa_0 = 0.4$ cm$^2$ g$^{-1}$, so that $f_{\rm Edd} = 0.5$ and $r_{\rm B} = 4 \times 10^{12}$ cm. The spherical domain has inner radius $0.2~r_{\rm B}$ and outer radius $6 r_{\rm B}$. We initialize the gas density and velocity according to the analytic solution as described in \citet{Krumholz2007}, in a manner such that each of our Lagrangian zones contains roughly equal mass. 

We enforce inflow boundary conditions by removing the innermost Lagrangian zone from the calculation when its outer radius drops below $0.25 r_B$. We then simultaneously add a zone at the outer edge of the computational domain with density equal to $\rho_{\infty}$ and with outer velocity equal to the velocity of the formerly outermost zone. As in the previous test, to compute the fluid pressure gradient at the outer boundary we linearly extrpolate the pressure from the outermost two zones. Radiation escapes through the outer boundary.

\begin{figure}
\includegraphics[width=0.5\textwidth]{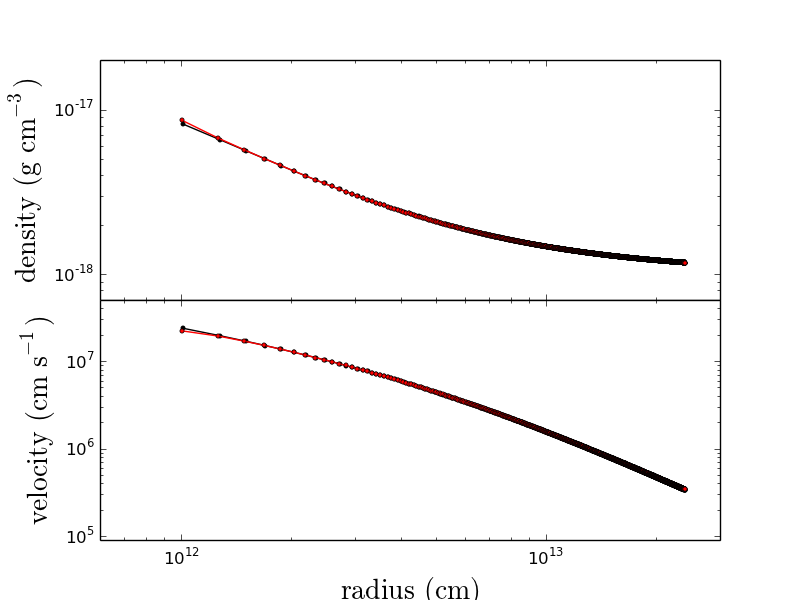}
\caption{The velocity and density profiles of a radiatively-inhibited Bondi accretion test, set up to replicate the corresponding test in  \citet{Krumholz2007} at time $t = 9.7 \, \, r_B/c_s$. The red curve is our computed solution and the black is the analytic solution. The maximum disagreement between these two solutions is $7$\% for the velocity in the innermost zone. The numerical parameters used in this test are listed in Table~\ref{Tab:NumericalParameters}.}
\label{Fig:BondiDensityAndVelocity}
\end{figure}

Figure \ref{Fig:BondiDensityAndVelocity} displays the gas density and velocity as a function of position at time $t =  9.7 \, \, r_B / c_s$. The fractional deviation between the computed and expected solutions is at most 7\% in the innermost zone. The average accretion rate over this time was 1.06 times the expected mass accretion rate computed from equation \ref{Eq:BondiRate}. 

\subsection{Steady sub-critical and super-critical radiating shocks}

A more complicated set of tests involve steady radiating shocks\footnote{We use the term ``shock'' here in a broad sense that also includes the case of very high upstream Mach number (e.g. our ${\cal M} = 70$ case) in which there is no embedded \emph{viscous} shock, although there is still a radiation-mediated shock.}. The structure of these shocks differs from the pure hydrodynamic case because radiation emitted by the shocked gas leaks out ahead and behind the shock, heating the gas and forming a radiative precursor region (upstream) and a radiative relaxation region (downstream). Additionally, the shock obeys a modified set of jump conditions in which the total energy and momentum carried by both gas and radiation is conserved \citep{Zeldovich1969}.

In the gray nonequilibrium diffusion approximation, there exists a semi-analytic solution for the shock structure \citep{Lowrie2008}. For the case where scattering is neglected, and for an adiabatic equation of state with fixed index $\gamma_{\rm ad}$ and mean particle mass $\mu m_p$, this solution is completely characterized by four dimensionless parameters: the Mach number,  ${\cal M}$, of the upstream gas in the rest frame of the shock, the ratio of the speed of light to the upstream sound speed ${\mathbb C}$, the ratio of the upstream radiation pressure (times 3) to upstream gas pressure ${\mathbb P}$, and the optical depth to the radiation, $\tau$, for a chosen comoving radiative extinction $\chi_0$ (units of cm$^{-1}$) and lab frame length scale $L$

\begin{eqnarray}
{\cal M} &=& v_u / a_u = v_u \sqrt{\frac{\mu m_p}{\gamma_{\rm ad} k_B T_{u,g}}} \\
{\mathbb P} &=& \frac{a_r T_{u,g}^4}{\rho_u a_u^2} = \frac{a T_{u,g}^3 \mu m_p}{\rho_u \gamma_{\rm ad} k_B } \\
{\mathbb C} &=& c / a_u = c \sqrt{\frac{\mu m_p}{\gamma_{\rm ad} k_B T_{u,g}}} \\
{\tau}  &=& \chi_0 L \, \, .
\end{eqnarray}

Here, quantities with subscript $u$ refer to upstream values\footnote{\citet{Lowrie2008} use a slightly different set of nondimensional parameters, but they are directly mappable to the ones listed here.}. For consistency with equations \ref{Eq:MassConservation} through \ref{Eq:EnergyConservation} we take $\rho_u$ and $v_u$ to be measured in the lab frame and $T_u$, $a_u$, and $\chi_0$ to be measured in the comoving frame, although this distinction is not made in \citet{Lowrie2008}. Also note that when setting these parameters, the upstream gas is considered to be in radiative equilibrium so that $T_{u,g}$ = $T_{u,r}$.

Following \citet{Lowrie2008} and \citet{Jiang2012}, we choose ${\mathbb P} = 10^{-4}$, $\mathbb{C} = 1.732 \times 10^3$, and $\tau = 577$. We take $L = 1$ cm, so that\footnote{In terms of the parameters used in \citet{Lowrie2008}, we are using $\sigma_a = 10^6$ and $\kappa = 1$.} $\chi_0 = 577$ cm$^{-1}$. We have also set $\mu = 0.5$. The upstream density, temperature, and velocity of the fluid can be determined from these values, and the downstream values can be calculated using the jump conditions and the procedure for solving them outlined in \citet{Bouquet2000}. 

We used the Eulerian version of the code. For ${\cal M} = 2$ and ${\cal M} = 5$, we initialized the computational domain with a step function obeying the jump conditions, not the full semi-analytic solution, and let the shock structure develop on its own. Then, once a structure emerged that was stable over multiple shock crossing times, we spatially translated this solution to compare the numerical shock structure to the semi-analytic solution. We used Dirichlet boundary conditions for the hydrodynamics solver. On the upstream side of the domain, we used reflecting boundary conditions for the radiation. On the downstream side, we let the radiation escape freely.

\begin{figure*}
\includegraphics[width=1.\textwidth]{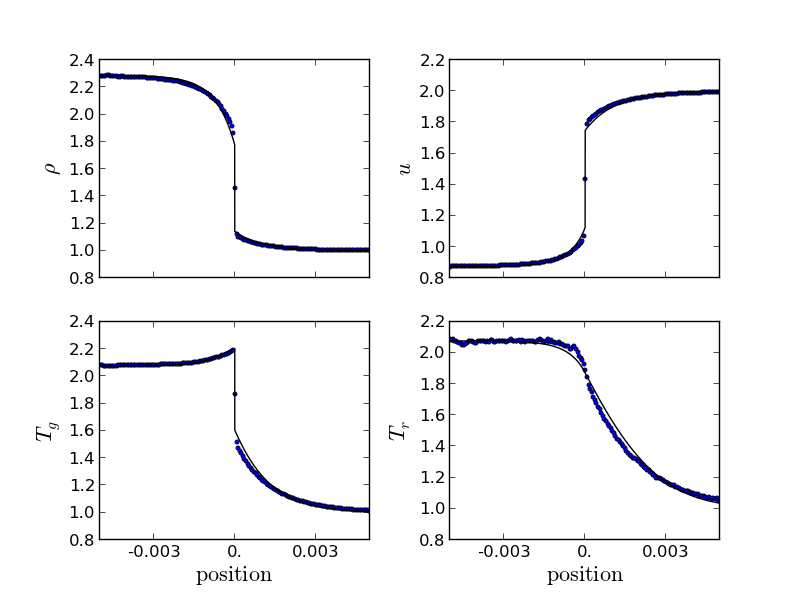} 
\caption{Steady radiating shock test as in \citet{Lowrie2008}, with ${\cal M} = 2$. The points are output from our Monte Carlo rad-hydro calculation, and the solid line is the semi-analytic solution. All hydrodynamic variables have been nondimensionalized (see text for details). The numerical parameters used for this test are listed in Table~\ref{Tab:NumericalParameters}.}
\label{Fig:LEM2}
\end{figure*}

\begin{figure*}
\includegraphics[width=1.\textwidth]{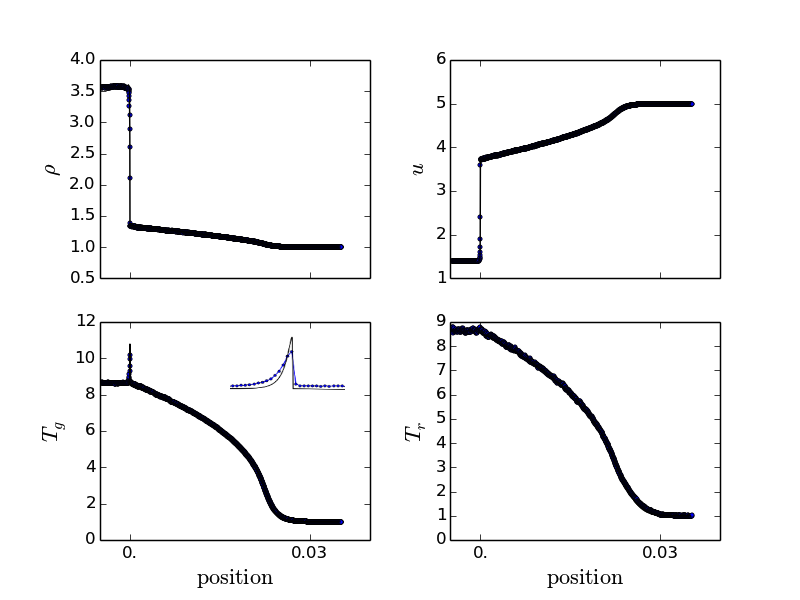} 
\caption{Similar to Figure \ref{Fig:LEM2}, but for ${\cal M} = 5$. The inset in the gas temperature plot is a zoomed-in plot in the region of the Zeldovich spike, and is not to scale with the rest of the figure.}
\label{Fig:LEM5}
\end{figure*}

Figure \ref{Fig:LEM2} shows the results for the ${\cal M} = 2$ case at $t = 1.0 \times 10^{-9}$ s, and Figure \ref{Fig:LEM5} shows the results for the ${\cal M} = 5$ case at $t = 1.9 \times 10^{-9}$ s. In general, there is excellent agreement with the semi-analytic solution. One slight issue relates to resolving the narrow Zeldovich temperature spike. For the ${\cal M} = 5$ case, we increased our resolution all the way to 2048 zones, and even then the spike is slightly underestimated. 

\begin{figure}
\includegraphics[width=0.5\textwidth]{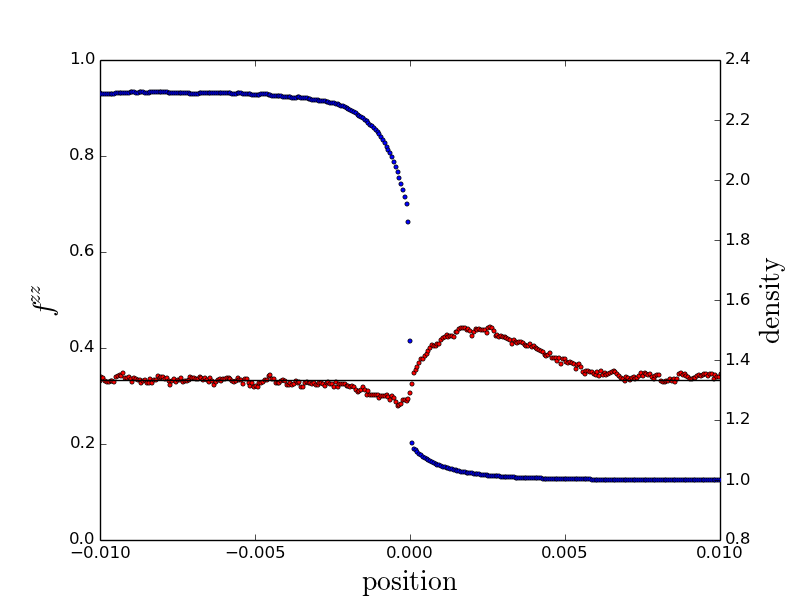} 
\caption{Eddington tensor element $f^{zz}$ (red) for the ${\cal M} = 2$ steady radiating shock test, with the nondimensional density (blue) over-plotted. The solid black line represents a constant value of $1/3$, which holds in the diffusion approximation. We that $f^{zz}$ does indeed drop below 1/3 near the shock, as previous authors have observed (see text for details).}
\label{Fig:EddingtonFactor}
\end{figure}

For the ${\cal M} = 2$ case, we also display our computed value of the Eddington tensor element $f^{zz}$ as a function of position in Figure \ref{Fig:EddingtonFactor}. FLD assumes that the diagonal elements of $f^{ij}$ never drop below $1/3$. We find, as in \citet{Sincell1999} and \citet{Jiang2012}, that $f^{zz}$ does indeed drop below 1/3 near the shock.

We also considered a stronger shock, with ${\cal M} = 70$. In this case, we initialized the problem with the steady-state solution and tested to make sure it maintained that solution over several shock crossing times. Also for the ${\cal M} = 70$ case, rather than implementing a constant radiative flux boundary condition on the downstream side, we used a reflecting boundary condition for the downstream radiation, and extended the downstream domain so that any spurious effects from this boundary condition did not have time to reach the region of interest near the shock.

\begin{figure*}
\includegraphics[width=1.\textwidth]{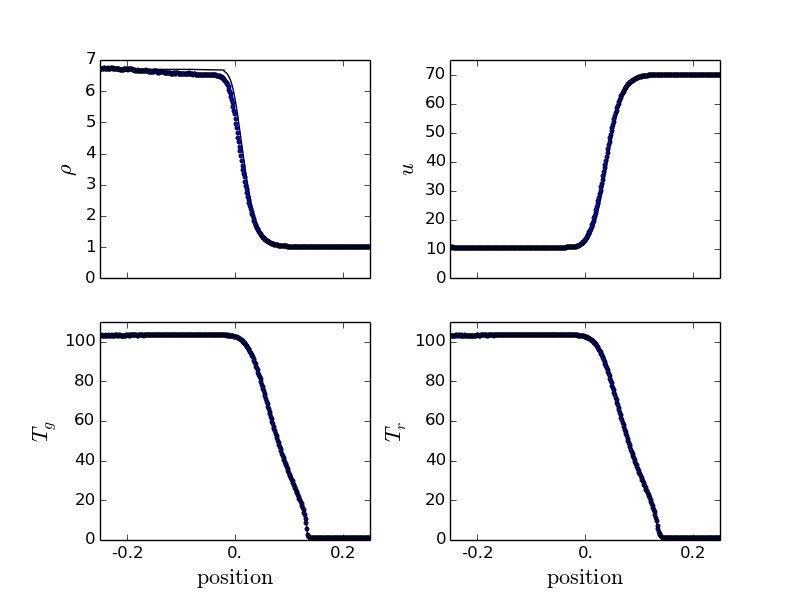} 
\caption{Similar to Figures \ref{Fig:LEM2} and \ref{Fig:LEM5}, but for ${\cal M} = 70$. }
\label{Fig:LEM70}
\end{figure*}

Figure \ref{Fig:LEM70} shows the results for the ${\cal M} = 70$ case at $t = 10^{-9}$ s. This case is particularly interesting because downstream of the shock, the radiation energy density is larger than the gas thermal energy by a factor of about 9. Here radiation pressure becomes dynamically important, and we are testing the behavior of the radiation pressure force in our code in the optically thick regime. We find good agreement between our computed results and the semi-analytic solutions for the gas and radiation temperatures. However, we find that the gas density on the downstream side is about 5\% lower than expected based on the jump conditions. The cause of this discrepancy is unclear, but it might be related to our method for coupling the radiation momentum source terms to the Godunov solver. Our hydro time-steps are large enough that we are in the highly implicit regime for the Monte Carlo, although our solution does not appear to change signficantly if we reduce our CFL number by a factor of 2.

\subsection{Non-steady radiating shocks}
\label{Sec:EnsmanShocks}

 This test involves a super-critical radiative shock driven by the supersonic motion of a piston into initially uniform and static gas, as defined by  \citet{Ensman1994}. The test has been revisited many times, including in \citet{Hayes2006}, in which the ZEUS-MP2 code was used to solve the problem while making use of the flux-limited diffusion approximation for the radiation. More recently, \citet{Noebauer2012} compared the results of their Monte Carlo radiation-hydrodynamics code to the ZEUS results for this problem. 

 The numerical parameters for these tests are reported in Table~\ref{Tab:NumericalParameters}. Additionally, we used reflecting boundary conditions for both the radiation and the hydrodynamics at the piston boundary. On the other side of the domain, we let radiation escape freely, and we used Dirichlet boundary conditions for the hydrodynamics.

\begin{figure}
\includegraphics[width=0.5\textwidth]{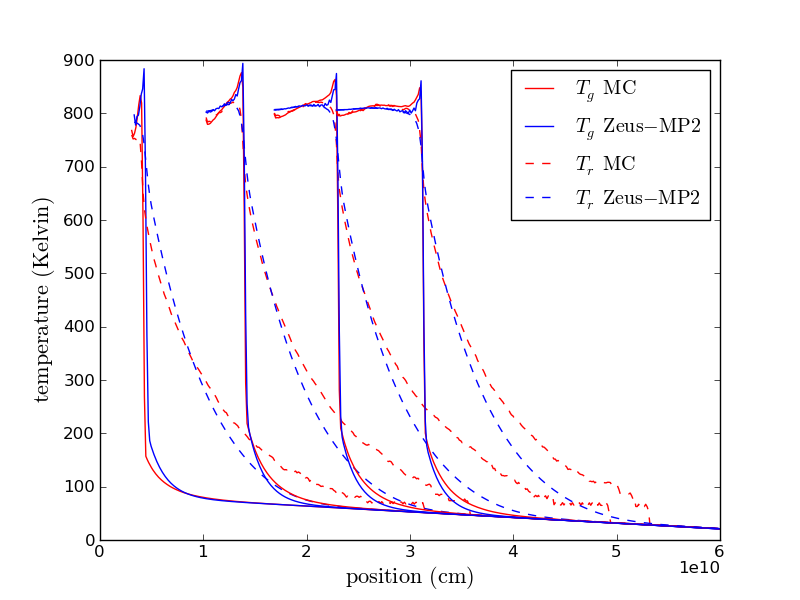}
\caption{Gas and radiation temperatures for the sub-critical moving shock test as described in \citet{Ensman1994}. We compare our solutions for the radiation and gas temperatures to those computed by the ZEUS-MP2 code. While the two calculations agree very well in the vicinity of the shock, the radiative precursor in the Monte Carlo calculation extends farther into the upstream gas, as was also observed in \citet{Noebauer2012}. The numerical parameters used in this test are listed in Table~\ref{Tab:NumericalParameters}.}
\label{Fig:EnsmanSub-Critical}
\end{figure}

\begin{figure}
\includegraphics[width=0.5\textwidth]{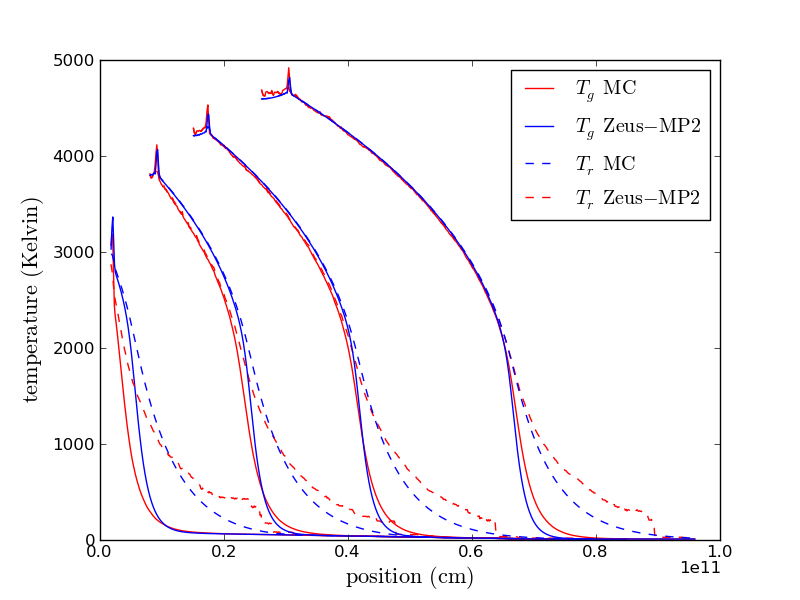}
\caption{Similar to Figure \ref{Fig:EnsmanSub-Critical}, but for a super-critical radiating shock.}
\label{Fig:EnsmanSuper-Critical}
\end{figure}

 In Figures \ref{Fig:EnsmanSub-Critical} and \ref{Fig:EnsmanSuper-Critical} we display our results for the sub-critical and super-critical versions of the test, respectively. The agreement between the two codes is encouraging. As \citet{Noebauer2012} found, we see deeper penetration of the radiation into the radiative precursor than in the FLD result. We also find that the radiative precursor in our results takes slightly more time to develop than in Zeus, and this might be due to our implicit treatment of the radiative cooling. 

\section{Radiation Force Calculation Using the Divergence of the Eddington Tensor}
\label{Sec:RadiationForceComparsion}

One of the main concerns with applying MCRT methods to RHD problems is that the estimators of the radiation field possess stochastic errors that may propagate into the dynamics. In general, the radiation force is more poorly sampled than the radiation energy deposition, due to the fact that packets traveling in opposite directions cancel out in the estimator of the flux. The problem becomes more acute in regions of high optical depth, where the radiation becomes nearly isotropic and the flux constitutes only a small fraction of the total radiation mean intensity. 

In this case, a better approach (mentioned in section \ref{Sec:Equations}) may be to use the divergence of the Eddington tensor (equation \ref{Eq:PressureDivergence}) to derive the radiation force. As noted before, this approach is only guaranteed to be accurate when the radiation is diffusing, but that is precisely the situation in which such an approach becomes most attractive. The $P^{zz}$ element of the radiation pressure tensor does not suffer from the same packet cancellation as does the flux, and so is typically better estimated. To calculate the radiation force, we used a simple centered difference to take a second-order spatial derivative of the pressure tensor. We note that there are more sophisticated methods for taking numerical derivatives of noisy data, and that these may lead to superior results.

\begin{figure}
\includegraphics[width=0.5\textwidth]{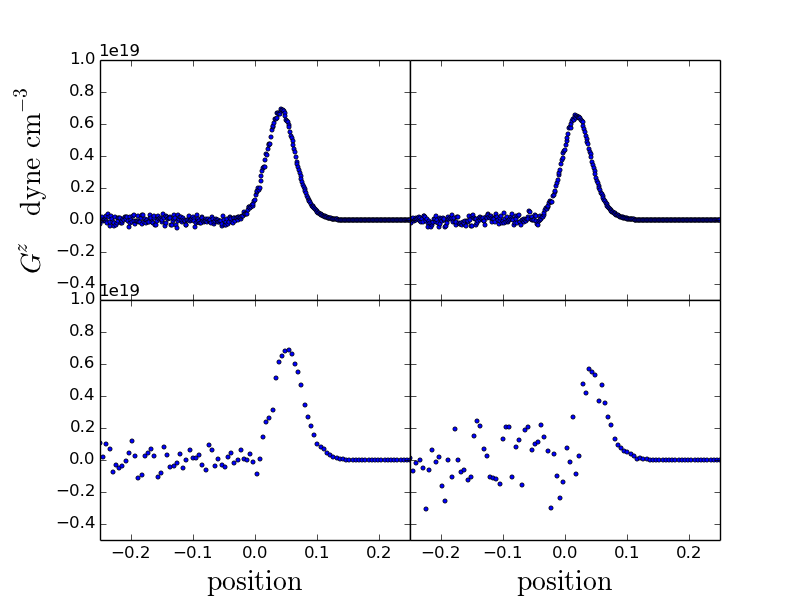}
\caption{Comparison of calculations of the radiation force in the ${\cal M}=70$ steady radiating shock problem. The left panels use the divergence of the pressure tensor to calculate the radiation force, whereas the right panels use the direct Monte Carlo summation method. The top two panels use numerical parameters as listed in Table~\ref{Tab:NumericalParameters}. The bottom two panels use a spatial resolution that is four times as coarse, and a maximum of 4000 packets per zone instead of 7680.}
\label{Fig:RadforceComparison}
\end{figure}

Figure \ref{Fig:RadforceComparison} compares the noise in the calculation of the radiation force in the ${\cal M} = 70$ steady radiating shock test using our two different methods (in all other tests, only the direct Monte Carlo estimator of the force was used, not the divergence of the pressure tensor). The figure demonstrates that although the two methods converge to a similar result at high spatial resolution and for a large number of packets, the pressure tensor divergence method converges must faster - it provides much less noise than the direct Monte Carlo summation method for coarser spatial resolutions and lower packet number. 

These results suggest the possibility of using MCRT in a hybrid approach with other radiation transport schemes. In particular, solution of the radiation moment equations require a closure relation, which is often taken to be an approximate analytic prescription. Solution of the MCRT, however,  provides estimator of the true Eddington tensor, which could then be used as a closure to the moment equations. In this case, the MCRT may not need to be run every time-step, allowing for a reduced computational load.

\section{Performance}
\label{Sec:Performance}

The relative performance of the MCRT compared to traditional radiation-hydrodynamics schemes depends sensitively on the particular problem at hand --- the spatial resolution, optical depth, degree of radiation domination, and level of tolerable noise. For the test problems discussed in the last section, we find that MCRT execution times are in some cases comparable to grey FLD techniques, and in others considerably more expensive.

Table \ref{Tab:Performance} summarizes the execution times for the Ensman super-critical shock test for a varying number of Monte Carlo packets employed, and two separate spatial resolutions. These tests were performed on a 2012 MacBook Pro laptop (2.6 Ghz Intel Core i7 processor) and compiled with g++. For comparison, we have included tests run with the flux-limited diffusion code Zeus-MP, run on the same machine and compiled with gfortran.

\begin{figure}
\includegraphics[width=0.5\textwidth]{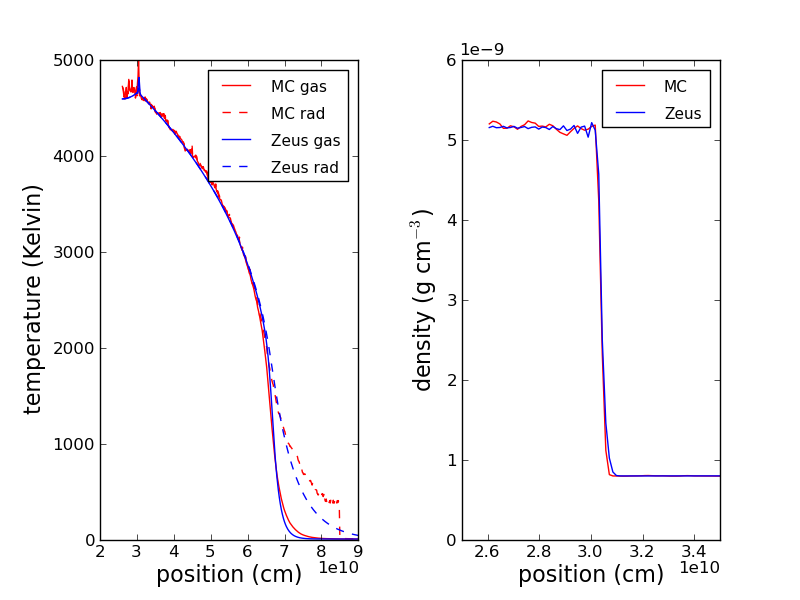}
\caption{The left panel corresponds to the fourth time output in Figure \ref{Fig:EnsmanSuper-Critical}, but with a maximum of only 100 packets emitted per zone per time-step, instead of 1000. The right panel shows the corresponding gas densities.}
\label{Fig:EnsmanSuper-CriticalFewerPackets}
\end{figure}

When a smaller number of packets is used, the radiation field in the MCRT calculation naturally possesses increased noise, as illustrated in Figure~\ref{Fig:EnsmanSuper-CriticalFewerPackets}. The error is most apparent in the high-temperature shocked gas, and in the stair step behavior at the leading edge of the radiative precursor. The latter effect arises because only a small number of high energy packets manage to diffuse ahead of the shock in any given time-step. This behavior is in part due to our unoptimized choice to emit equal numbers of packets in every zone, despite the fact that the emissivity behind the shock is at least $10^6$ times greater than that of the coldest gas ahead of it. Applications of so-called importance sampling techniques may substantially reduce the error without increasing the execution time. In particular, one could increase the number of high-energy packets emitted near the shock interface, while at the same time reducing the number of low-energy packets emitted in the pre-shock region. 

It is comforting to see that, despite the noisy radiation field of Figure~\ref{Fig:EnsmanSuper-CriticalFewerPackets}, the gas density suffers less from noise. This is because the gas properties are determined by the radiative heating and acceleration integrated over many time-steps. These time-averaged quantities are  more accurately sampled than the instantaneous  radiation field snapshot plotted in the figure. In general, we find that for problems where gas energy dominates, the dynamics of the problem are rather robust against instantaneous radiation noise. For problems where radiation energy and pressure dominate, the radiation noise is more problematic and can propagate into the gas properties. The deleterious effect of noise may also be more significant in multi-dimensional simulations where instabilities might develop. 

\begin{figure}
\includegraphics[width=0.5\textwidth]{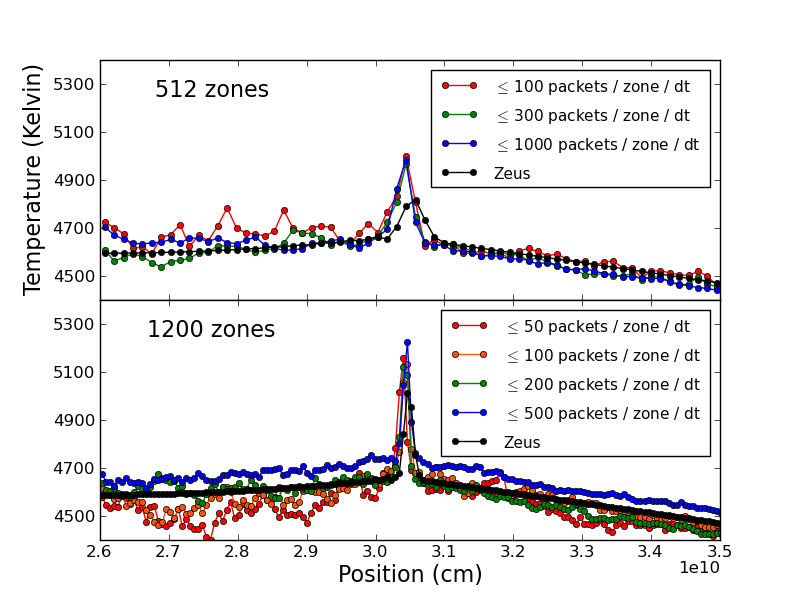}
\caption{Gas temperatures calculated for the Ensman super-critical shock test, zoomed in to the region surrounding the Zeldovich spike, for various packet counts and two separate spatial resolutions.}
\label{Fig:NoiseComparison}
\end{figure}

Figure~\ref{Fig:NoiseComparison} shows the gas temperatures computed for each of the calculations in Table~\ref{Tab:Performance}, zoomed in to the region surrounding the Zeldovich spike. In addition to the noise present at the scale of a few zone widths, the value of the temperature averaged over larger scales also varies between the individual Monte Carlo calculations at the level of a few percent, as is evident in Figure~\ref{Fig:NoiseComparison} and listed in Table~\ref{Tab:Performance}. In order to quantify the small-scale noise, we focus on the region upstream (left) of the spike, where the effect of the noise is most severe. We apply an offset to the Monte Carlo temperatures so that their mean value in this region matches that of the Zeus calculation. Then, we measure the root mean square difference between these offset gas temperatures and the temperatures computed by Zeus, excluding the 3 zones directly adjacent to the left boundary. The effect of increasing the number of packets on the RMS error for the four calculations with 1200 zones agrees especially well with the rule of thumb that the random error should scale as $1/\sqrt{N}$ where $N$ is the number of packets employed. We see that to decrease the RMS error associated with small-scale noise to within 1\%, the CPU time requirement is approximately four times that of Zeus in the runs employing 512 zones, and twice that of Zeus in the runs employing 1200 zones. 

\begin{deluxetable*}{cccc}
\tablecaption{Performance comparison for the Ensman super-critical shock test}
\tablehead{\colhead{Description} & \colhead{Mean gas} &  \colhead{RMS noise (K),} & \colhead{CPU Time} \\ & temperature (K) & percent error & (minutes) }
\startdata
Zeus 512 zones & 4617.5  &  & 1.8 \\
MC 512 zones,  100 packets & 4686.9 & 44.7 (0.95~\%) & 6.4 \\
MC 512 zones,  300 packets & 4618.0 & 29.3 (0.63~\%) & 18.3 \\
MC 512 zones, 1000 packets & 4639.0 & 24.7 (0.53~\%) & 56.4 \\
 &  &  &  \\
Zeus 1200 zones & 4613.8  & &  17.8\\
MC 1200 zones,  50 packets & 4535.5 & 50.1 (1.1  \%) & 35.4  \\
MC 1200 zones, 100 packets & 4574.1 & 35.5 (0.75 \%) & 68.3  \\
MC 1200 zones, 200 packets & 4622.2 & 25.2 (0.55 \%) & 133.6 \\
MC 1200 zones, 500 packets & 4680.6 & 14.3 (0.31 \%) & 298.3 \\

\enddata
\tablecomments{The number of packets in the description refers to the maximum number of packets emitted per zone per time step. All other numerical parameters are as listed in Table~\ref{Tab:NumericalParameters}. The mean gas temperature is computed in the upstream region left of the Zeldovich spike, excluding the three zones nearest to the left boundary. For the details of how the RMS noise is computed, please see the text. }
\label{Tab:Performance}
\end{deluxetable*}

The potential performance advantages of the MCRT method would become more apparent if, instead of comparing to Zeus-MP, we were to compare to a  non-grey radiation  code. For most deterministic transport methods, the execution time scales with the number of angle bins and frequency groups employed, and therefore become significantly more expensive than grey FLD. Our MCRT calculations, on the other hand, already include the angular information and can be run in multi-frequency mode with minimal additional computational expense. Photon packets are distributed across the relevant frequency range and, because the radiation force four-vector is given by integrals over frequency, no additional packets are needed to construct  source term estimators of comparable noise, at least in the case that the opacity has a reasonably smooth frequency dependence. In cases where the opacity has sharp dependencies (e.g., lines),  importance sampling technique can be used to increase packet statistics at the most important frequencies. Convergence tests  varying the number of packets  would be required to determine whether the frequency sampling had been sufficient.

As already mentioned, the execution time of the MCRT code is highly problem-dependent, in particular because of the well-known inefficiency of Monte Carlo methods in regions of high optical depth, where many photon interactions must be tracked per time-step. For high optical depth cases, a substantial speed-up can be obtained through the inclusion of the discrete diffusion technique, which has been described for the gray radiation case in \citet{Gentile2001, Densmore2007} and the non-gray case in \citet{Abdikamalov2012}.

For problems with one spatial dimension, or problems of higher dimension and sufficiently coarse spatial resolution, the entire computational domain can be stored in the memory of a single computational node. With the added fact that Monte Carlo packets propagate independently of one another over a single time-step, this permits an ``embarrassingly parallelizable'' treatment for the radiation portion of the problem. The transport step may be replicated over as many computational nodes as are available, and then the results of the packet propagation during each time-step for each node can be summed together with an Message Passing Interface (MPI) reduction. 

\begin{figure}
\includegraphics[width=0.5\textwidth]{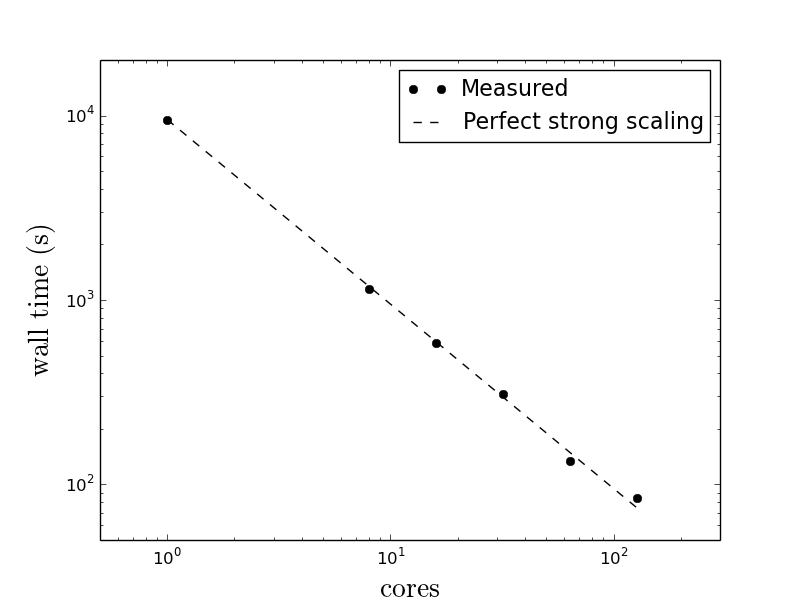}
\caption{Parallel scaling for the Ensman sub-critical shock test.}
\label{Fig:ParallelScaling}
\end{figure}

We have run MPI-parallelized versions of the Ensman sub-critical shock test in which the total number of photon packets per time-step is held constant, but is divided over varying numbers of CPUs. Although we have only parallelized the radiation portion of the code, the CPU time required to execute the hydro update is negligible compared to the radiation. We see perfect strong parallel scaling in the time for this test, which is to say that the amount of wall time needed to complete the test is cut in half each we double the number of cores we use, as shown in Figure \ref{Fig:ParallelScaling}. 

\section{Conclusions}
\label{Sec:Conclusion}

We have demonstrated that MCRT coupled to both Lagrangian and Eulerian hydrodynamics solvers can result in accurate, robust treatments of RHD problems, including those in which the radiation energy dominates. Although we have focused here on 1-dimensional test problems,  our Eulerian code is multi-dimensional, and subsequent studies will address astrophysical problems in higher spatial dimensions.

Our approach makes use of the implicit MCRT method to allow us to take hydrodynamical time-steps much larger than the gas cooling time. We also showed how to use Monte Carlo estimators to construct expressions for the radiation force four-vector $G^i$ that are accurate to all orders of $v/c$, although the hydrodynamics equations are only solved to order $v/c$. We compared simulations using our exact expression for $G^i$ to those using a more approximate expression based on the divergence of the radiation pressure tensor, which is valid when the radiation is in the diffusion regime. We found that the latter method can lead to a significant reduction in Monte Carlo noise in cases of coarse spatial resolution. In most of the problems studied here, the presence of stochastic noise did not introduce substantial error in the dynamics, however the effects of noise become a larger concern in problems where radiation energy is strongly dominated.

Several additional refinements will be explored in the future. We will consider the use of a more sophisticated treatment of the radiative source terms in the Godunov scheme. Improvements in performance may be realized by incorporating the discrete diffusion technique. It is straightforward to incorporate the effects of more complicated radiation-matter interactions, including photoionization and anisotropic scattering processes such as Compton scattering with Klein-Nishina corrections. Possible applications of this technique include radiatively-launched winds from galaxies, tidal disruptions of stars, shock breakouts and ejecta-ISM interactions in supernovae.

\section{Acknowledgments}
We thank Phil Colella for guidance in the developing the Godunov solver.
We thank Weiqun Zhang for providing the script to compute the semi-analytic radiating shock solutions.
We thank Ulrich Noebauer for helpful correspondence regarding the comparison to Zeus-MP2.

Throughout the work, NR was supported by the Department of Energy Office of Science
Graduate Fellowship Program (DOE SCGF), made possible in part by the American Recovery
and Reinvestment Act of 2009, administered by ORISE-ORAU under contract no. DE-AC05-
06OR23100.

DK is supported by a Department of Energy Office
of Nuclear Physics Early Career Award (DE-SC0008067). 

This research used resources of the National Energy Research Scientific
Computing Center, which is supported by the Office of Science of the
U.S. Department of Energy under Contract No. DE-AC02-05CH11231. This
research used resources of the Oak Ridge Leadership Computing Facility
at the Oak Ridge National Laboratory, which is supported by the Office
of Science of the U.S. Department of Energy under Contract
No. DE-AC05-00OR22725.

\bibliographystyle{apj}

\end{document}